\documentclass[sn-chicago]{sn-jnl}

\usepackage{fix-cm}
\usepackage{graphicx}%
\usepackage{multirow}%
\usepackage{amsmath,amssymb,amsfonts}%
\usepackage{amsthm}%
\usepackage{mathrsfs}%
\usepackage[title]{appendix}%
\usepackage[dvipsnames]{xcolor}
\usepackage{textcomp}%
\usepackage{manyfoot}%
\usepackage{booktabs}%
\usepackage{algorithm}%
\usepackage{algorithmicx}%
\usepackage{algpseudocode}%
\usepackage{listings}%
\usepackage{longtable}%
\usepackage{pgfplots}
\usepackage{caption}
\usepackage{subcaption}
\usepackage{smartdiagram}
\usepackage{geometry}
\usepackage{multirow}
\usepackage{float}
\usepackage{subcaption}
\usepackage{amsmath}
\usepackage{graphicx}
\usepackage[colorinlistoftodos]{todonotes}
\usepackage{mathtools}
\usepackage{pgfplotstable,booktabs}
\usepackage{tikz}
\usetikzlibrary{shapes.geometric}
\usetikzlibrary{positioning}
\usetikzlibrary{trees,shapes.geometric,arrows,quotes}
\usetikzlibrary{fit,calc}
\usetikzlibrary{patterns}
\usetikzlibrary{quantikz}  
\usepackage{forest}
\useforestlibrary{edges}
\usepackage{tikz-cd} 
\usepackage{smartdiagram}
\usesmartdiagramlibrary{additions}
\usepackage[english]{babel}
\usepackage[utf8x]{inputenc}
\usepackage[T1]{fontenc}

\usepackage{soul}
\usepackage{xcolor}

\usepackage{framed}
\usepackage{lipsum}
\usepackage{lineno}

\usepackage{ifpdf}
\ifpdf
  \usepackage{hyperref}
\else
  \usepackage{hyperref}
  \usepackage{breakurl}
\fi
\usepackage{silence}
\theoremstyle{thmstyleone}%
\theoremstyle{thmstyletwo}%

\theoremstyle{thmstylethree}%

\begin{document}

\title[Quantum Software Engineering and Potential of Quantum Computing in Software Engineering Research: A Review]{Quantum Software Engineering and Potential of Quantum Computing in Software Engineering Research: A Review}


\author*[1]{\fnm{Ashis} \sur{Kumar Mandal}}\email{ashis.62@gmail.com}

\author[1]{\fnm{Md} \sur{Nadim}}\email{mdn769@usask.ca}

\author[1]{\fnm{Chanchal K.} \sur{Roy}}\email{chanchal.roy@usask.ca}

\author[1]{\fnm{Banani} \sur{Roy}}\email{banani.roy@usask.ca}

\author[1]{\fnm{Kevin A.} \sur{Schneider}}\email{kevin.schneider@usask.ca}

\affil[1]{\orgdiv{Computer Science}, \orgname{University of Saskatchewan}, \orgaddress{ \city{Saskatoon}, \state{Saskatchewan}, \country{Canada}}}


\abstract{

Research in software engineering is essential for improving development practices, leading to reliable and secure software. Leveraging the principles of quantum physics, quantum computing has emerged as a new computational paradigm that offers significant advantages over classical computing. As quantum computing progresses rapidly, its potential applications across various fields are becoming apparent. In software engineering, many tasks involve complex computations where quantum computers can greatly speed up the development process, leading to faster and more efficient solutions. With the growing use of quantum-based applications in different fields, quantum software engineering (QSE) has emerged as a discipline focused on designing, developing, and optimizing quantum software for diverse applications.
 This paper aims to review the role of quantum computing in software engineering research and the latest developments in QSE. To our knowledge, this is the first comprehensive review on this topic. We begin by introducing quantum computing, exploring its fundamental concepts, and discussing its potential applications in software engineering. We also examine various QSE techniques that expedite software development. Finally, we discuss the opportunities and challenges in quantum-driven software engineering and QSE. Our study reveals that quantum machine learning (QML) and quantum optimization have substantial potential to address classical software engineering tasks, though this area is still limited. Current QSE tools and techniques lack robustness and maturity, indicating a need for more focus. One of the main challenges is that quantum computing has yet to reach its full potential.}




\maketitle

\section{Introduction and Motivation}

Classical computing has seen remarkable progress over the past few decades, mainly due to advances in semiconductor technology. The principle of Moore's law, which suggests that computing power doubles roughly every 18 months, has been held for a long time \citep{schaller1997moore}. However, we are now nearing the boundaries of Moore's law, with computing power showing signs of levelling off. This necessitates the exploration of new technologies to enhance computational abilities further. About forty years ago, Richard Feynman proposed the concept of quantum computing, a machine that could harness the laws of quantum physics to create a more robust computational system \citep{feynman1985quantum}. Given the constraints of existing classical computational capabilities, quantum computing has emerged as a solution to many computationally intensive problems. In contrast to classical computers, which use bits (0s and 1s) for digital representation, quantum computing uses qubits. A qubit can exist not only in the 0 or 1 state but also in a superposition of both states. This characteristic allows quantum computers to perform computations on multiple states simultaneously. By leveraging quantum mechanical phenomena such as entanglement, superposition, and interference, quantum computers can process a multitude of possibilities simultaneously \citep{gill2022quantum}. The probabilistic nature of quantum computers and their capacity to exploit quantum parallelism offer the potential for exponential speedup in specific computational tasks, including optimization problems, machine learning, and simulation \citep{paudel2022quantum}. Despite the promise of quantum computers to exponentially increase computing power with more qubits, they are susceptible to decoherence, posing a significant challenge. This vulnerability necessitates intricate error correction codes, which add to the overhead and resource requirements \citep{preskill1998quantum}. Currently, quantum computing devices, characterized by a limited number of qubits and high error rates, are often classified as Noisy Intermediate-Scale Quantum (NISQ) devices \citep{bharti2022noisy}. Although building large-scale, fault-tolerant quantum computers remains a daunting task, NISQ devices already show their potential to tackle various complex real-world problems.

With their superior computational efficiency compared to the classical ones, quantum computing paradigms are being applied to solve problems in several fields such as drug discovery, material science, machine learning, artificial intelligence, cryptography, security, and optimization \citep{marella2020introduction}. In the realm of software engineering, numerous tasks that require acceleration could benefit from using quantum computing. For instance,  quantum machine learning (QML) could be utilized for software defect prediction, while quantum optimization algorithms could be employed to generate and minimize software test cases. In both cases, quantum computing speeds up the computation process. Moreover, the advent of innovative quantum-oriented software solutions has catalyzed the evolution of quantum software engineering (QSE) \citep{9340056}, which offers a framework of concepts, principles, and guidelines for the creation, maintenance, and evolution of quantum applications. Research into areas such as quantum bug mitigation, quantum code testing, and quantum software design is ongoing \citep{Dwivedi2024}, highlighting the potential for advancements in this field. Even though this new approach could speed up tasks in classical software engineering, the investigation of quantum computing in this field is still in its initial phases. The primary objective of this paper is to examine state-of-the-art techniques in QSE and to explore how quantum computers can speed up classical software engineering processes. The following research questions have been formulated for this review.

 \begin{itemize}
\item RQ1: In what ways have QML methods been applied to address software engineering tasks? How can a generic model be developed using QML to aid research in software engineering?
\item RQ2: How have software engineering tasks been tackled using quantum optimization and search algorithms? What could be a general framework for a quantum optimization-based approach to solve software engineering tasks?
\item RQ3: What progress has been made in the field of QSE, and what tools are available to develop the creation of quantum applications?
\item RQ4: What are the challenges and opportunities related to QSE and quantum-assisted software engineering?
 \end{itemize}

The structure of the paper is as follows. Section 2 presents an overview of quantum computing. The methodology for searching research articles is detailed in Section 3. Quantum computing in the context of software engineering research is explored in Section 4. The current state of QSE is outlined in Section 5, while Section 6 highlights the opportunities and challenges. Section 7 provides a discussion of the research questions, and Section 8 presents threats to validity. Finally, Section 9 concludes the paper.


\section{Background of Quantum Computing}

\begin{figure}[!htb]
\centering
	\begin{subfigure}[b]{0.4\textwidth}
		\centering
		\includegraphics[width=\textwidth]{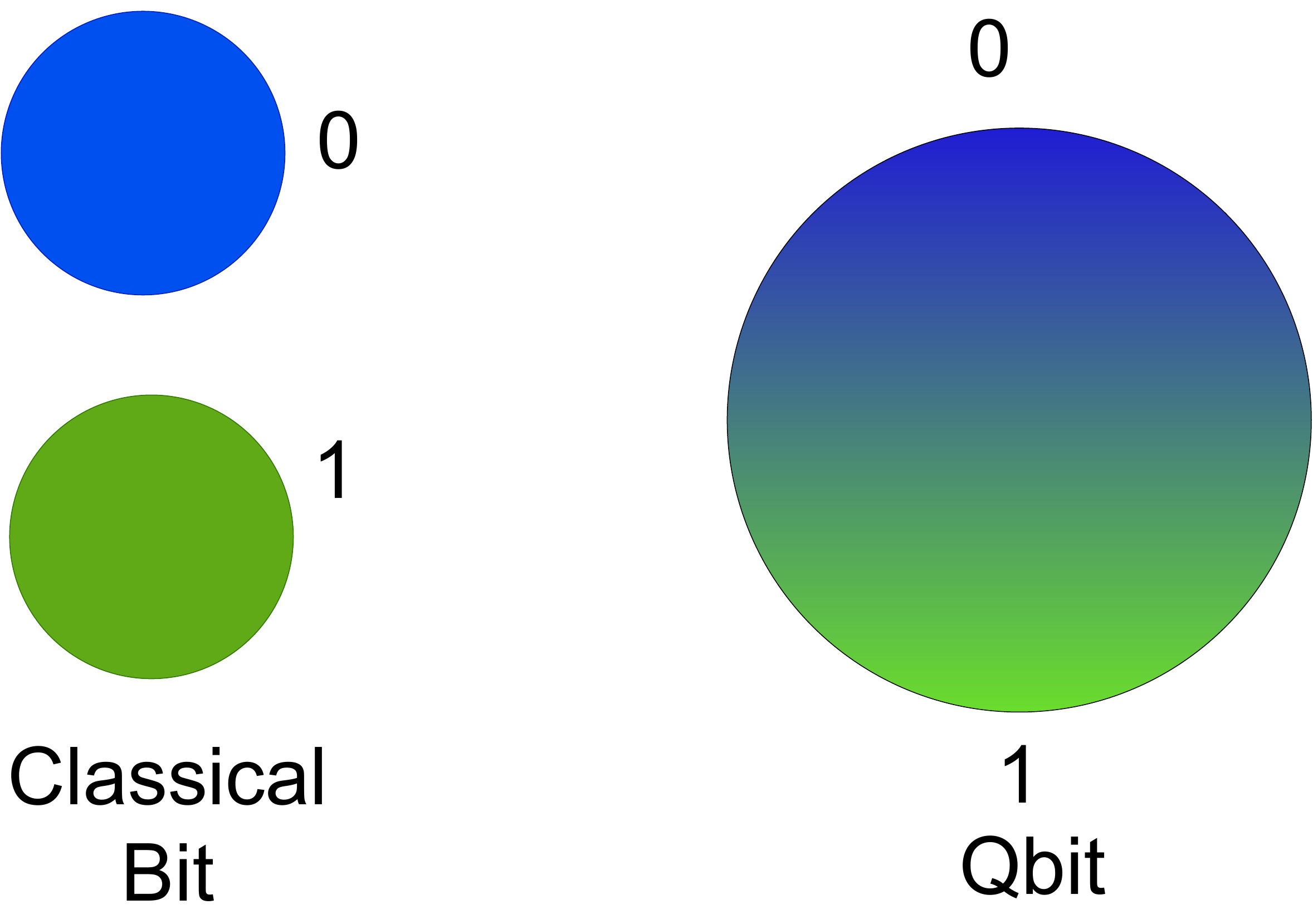}
		\caption{Superposition}
		\label{fig:superposition}
	\end{subfigure}
	\hfill
	\begin{subfigure}[b]{0.4\textwidth}
		\centering
		\includegraphics[width=\textwidth]{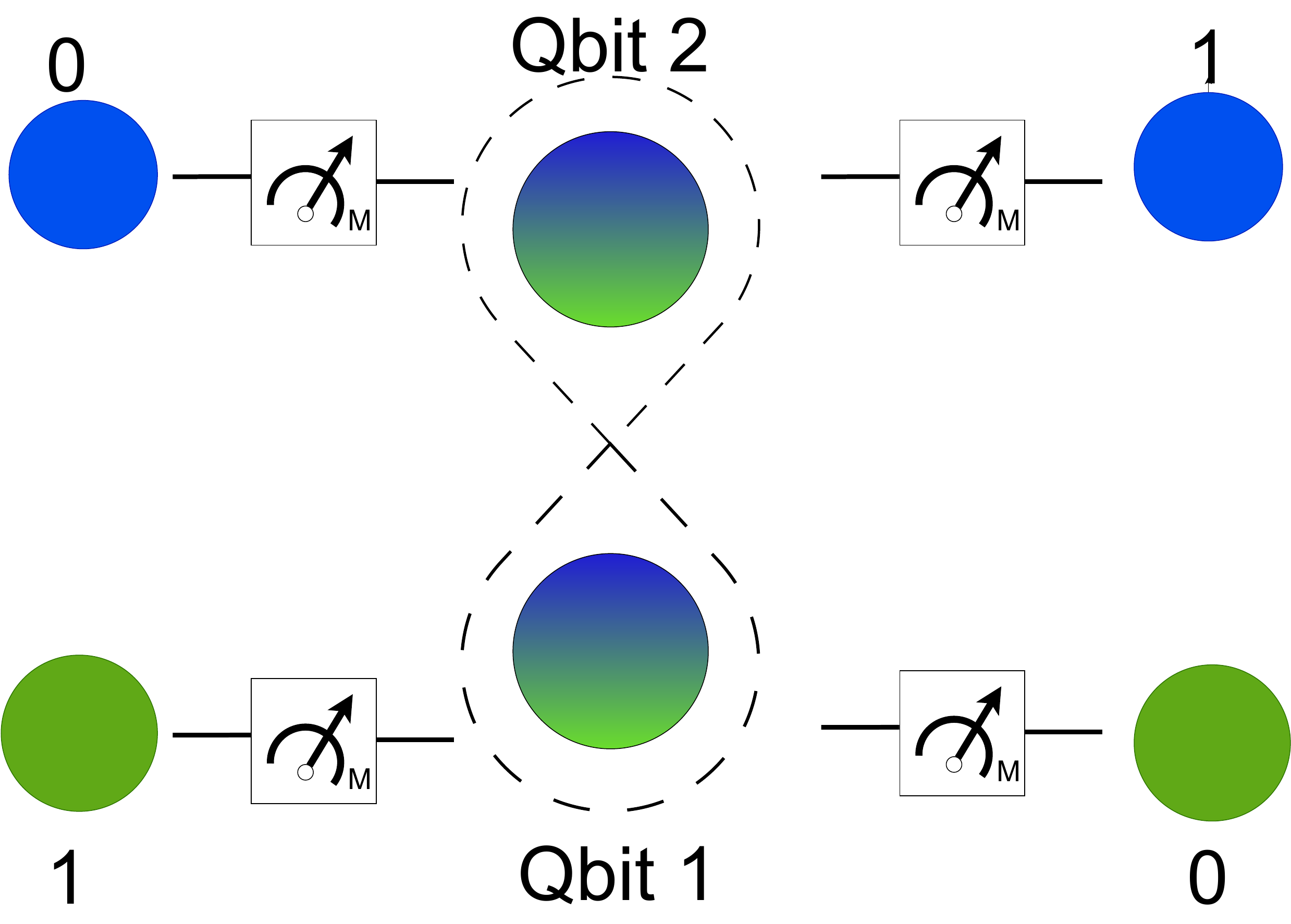}
		\caption{Entanglement}
		\label{fig:entanglement}
	\end{subfigure}
	\hfill
	\begin{subfigure}[b]{0.4\textwidth}
		\centering
		\includegraphics[width=\textwidth]{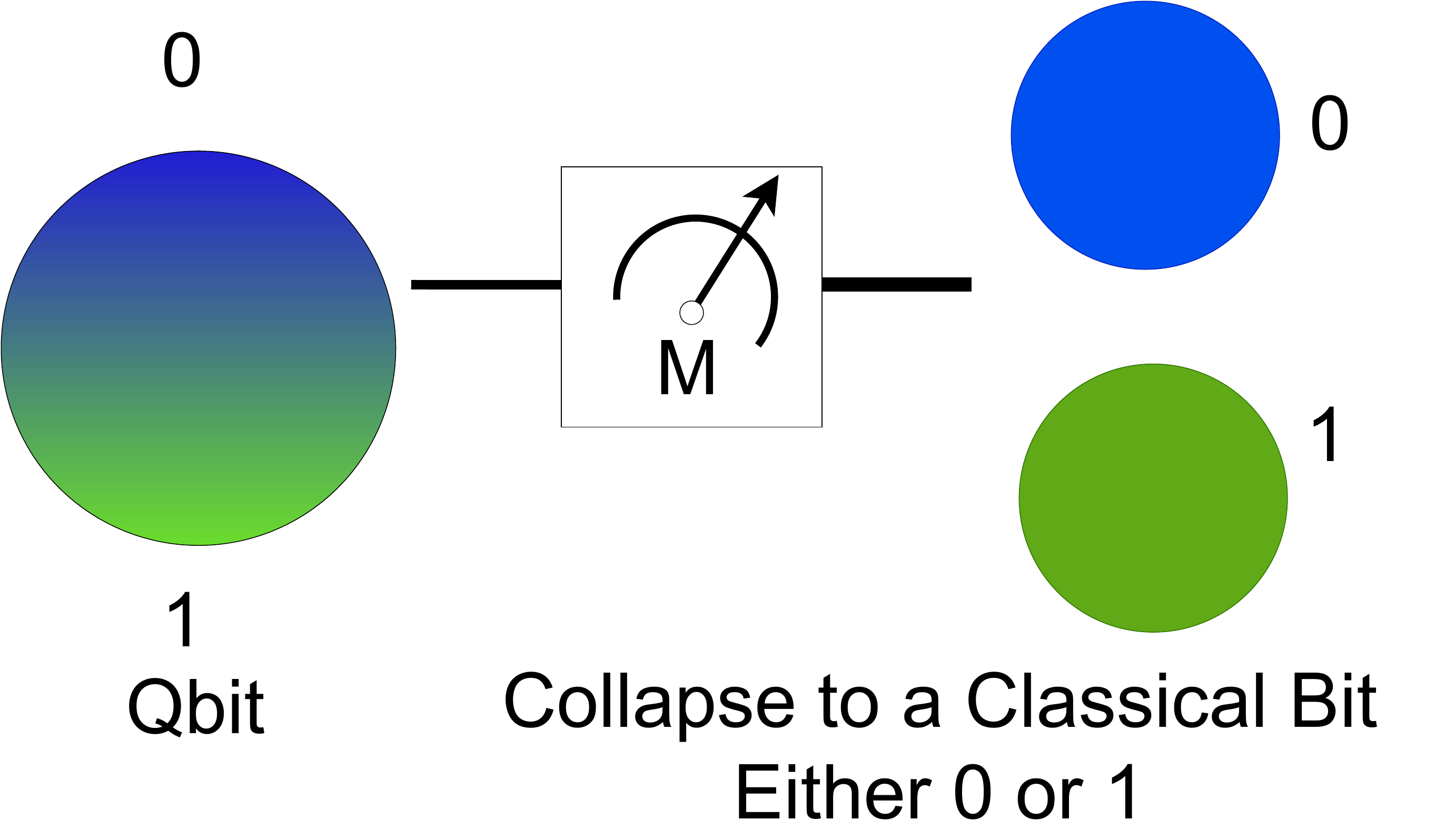}
		\caption{Measurement}
		\label{fig:measurement}
	\end{subfigure}
	\hfill
	\begin{subfigure}[b]{0.4\textwidth}
		\centering
		\includegraphics[width=\textwidth]{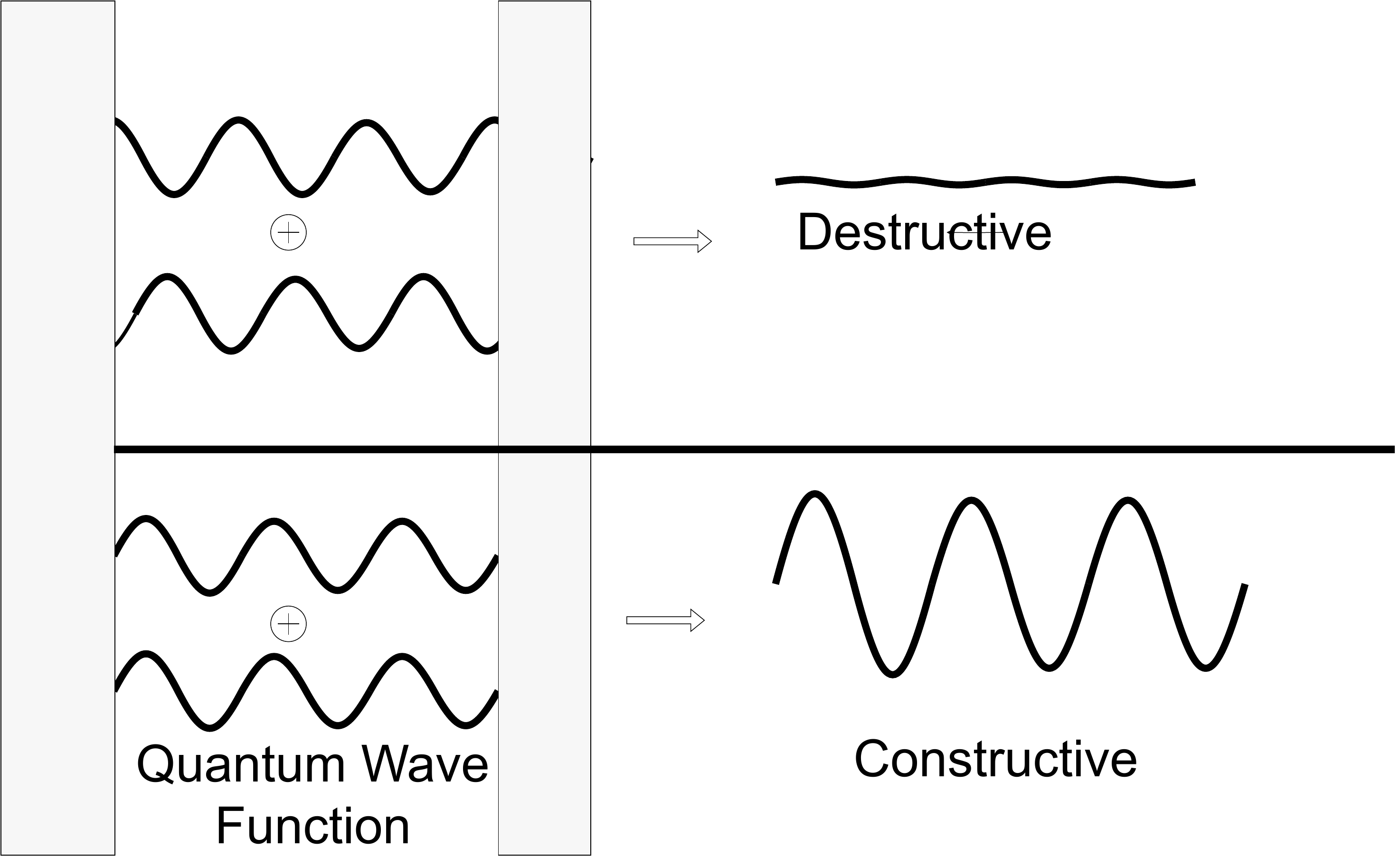}
		\caption{Interference}
		\label{fig:interference}
	\end{subfigure}
	\caption{Superposition, entanglement, measurement, and interference processes of a quantum computer.}
	\label{fig:threeQuantum}
\end{figure}

The principles of quantum mechanics, such as superposition, entanglement, interference, and measurement, are fundamental concepts in building quantum computers \citep{bouwmeester2000physics}.
Superposition \citep{kovachy2015quantum} is a fundamental property of quantum systems, allowing them to exist in multiple states at the same time. Unlike classical bits, which can only be 0 or 1, a qubit can represent both 0 and 1 simultaneously due to superposition.
Quantum entanglement is a crucial phenomenon in which two or more quantum particles remain interconnected regardless of distance \citep{Horodecki2009}.  That is, one particle's state instantaneously determines the state of the other(s), regardless of the physical distance separating them. In the case of two entangled qubits, measuring one qubit as zero immediately reveals that the other qubit is in  state one, even without directly measuring it.  Quantum interference \citep{Hornberger2012}, caused by overlapping wave functions, affects qubit states by altering measurement probabilities. Constructive interference boosts certain outcomes, while destructive interference cancels others, forming the basis of many quantum algorithms.  Quantum measurement involves collapsing a superposition of states into a single, definite classical state, either 0 or 1, through the process of measurement. During this collapse, the quantum properties of the system are lost. Figure \ref{fig:threeQuantum} illustrates the concepts of superposition, entanglement, interference, and measurement in quantum systems.

In the field of quantum computing, two major computational paradigms exist: the universal gate quantum model and the AQC model \citep{MCGEOCH2020169}. The universal gate quantum model is a versatile approach that addresses a wide range of problems. It does this by representing these problems through various quantum gates and then seeking the best solutions \citep{gyongyosi2020unsupervised}. This model is used by leading companies such as IBM, Google, Honeywell, Intel, and Rigetti. These companies harness the potential of the universal gate quantum model to tackle complex computational challenges. On the other hand, AQC leverages the unique properties of quantum computing, including superposition, entanglement, and tunnelling, to address a specific subset of optimization problems \citep{albash2018adiabatic}. D-Wave machines are prominent examples designed to implement the AQC model. Although both models utilize quantum mechanics, they differ in their operational methods \citep{mizuno2021research}. The gate model mirrors conventional digital computation, while the AQC model adopts a more continuous, analog-style approach \citep{mcgeoch2022adiabatic}.

\begin{figure}[!htb]
    \centering
	\tikzset{process/.style={draw, text width=8cm, minimum height=1.5cm, align=center, font=\rmfamily, rectangle,fill=CadetBlue!40,rounded corners=2pt,},
		arrow/.style={-latex}
	}
	\begin{tikzpicture}[node distance=2cm]

        \node (formulation) [process] {\textbf{Problem Formulation}: Define the problem using an SDK};
        \node (compilation) [process, below of=formulation] {\textbf{Compilation}: Translate algorithms into QPU instructions};
        \node (communication) [process, below of=compilation] {\textbf{Communication}: Send instructions to the QPU};
        \node (processing) [process, below of=communication] {\textbf{Quantum Processing}: Execute instructions using qubits};
        \node (measurement) [process, below of=processing] {\textbf{Measurement}: Collapse quantum states to classical outputs};
        \node (postprocessing) [process, below of=measurement] {\textbf{Post-Processing}: Process results on the classical system};
        
        \draw [arrow] (formulation) -- (compilation);
        \draw [arrow] (compilation) -- (communication);
        \draw [arrow] (communication) -- (processing);
        \draw [arrow] (processing) -- (measurement);
        \draw [arrow] (measurement) -- (postprocessing);
        
        \end{tikzpicture}
            \caption{ The workflow diagram for the typical quantum computing process}
            \label{fig:flowchart_qc}
\end{figure}
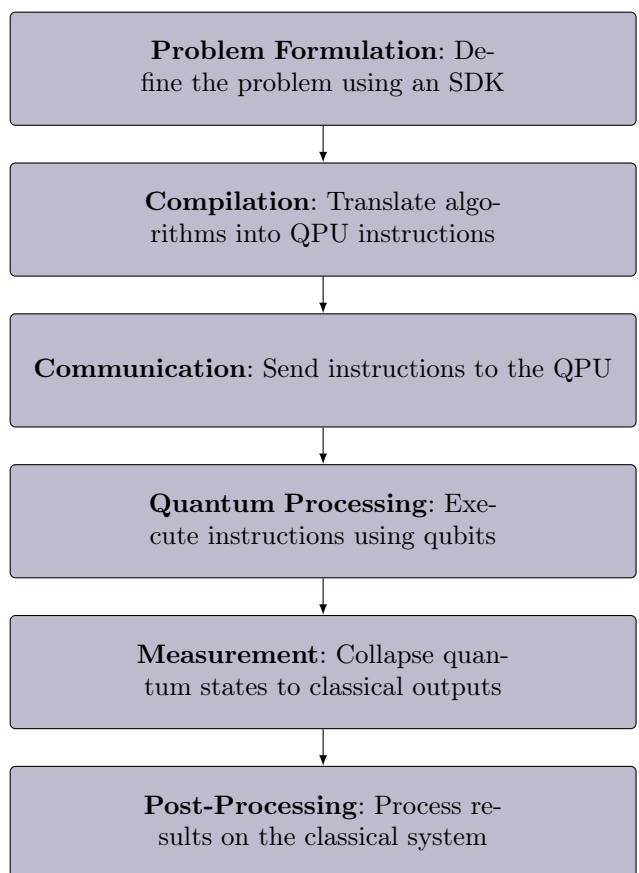

The core of a quantum computer is QPU, where quantum computations occur. QPUs contain qubits, the fundamental units of quantum information, and the electronics to control them. Currently, in the NISQ era, QPUs have limitations in the number of qubits and susceptibility to errors \citep{li2019tackling}. QPUs frequently work with classical computers in hybrid systems to avoid these limitations, acting more like specialized accelerators than complete replacements \citep{ding2022quantum}. To utilize QPUs for computation, the desired problems are first coded using Quantum Software Development Kits (SDKs). SDKs provide libraries, simulators, and tools for building quantum applications. The SDKs compile the code into instructions so the QPU can understand and manage the communication with the QPU, sending instructions, which are then processed in the QPU, and the final results are sent back to the classical computer. Figure \ref{fig:flowchart_qc} depicts the workflow diagram of a typical quantum computing process.

\subsection{Power of quantum computing with simple example}

\begin{figure}[!htb]
    \centering
\tikzset{
    mycircle/.style={fill=#1!80,draw=none, circle,minimum size=0.5cm},
    mycylinder/.style={cylinder, fill=CadetBlue!80, shape border rotate=180, aspect=0.25, minimum height=3cm, minimum width=0.5cm},
    myrectangle/.style={rounded corners, Gray, rectangle,fill=GreenYellow,fill opacity=0.50},
}
\begin{tikzpicture}[scale=0.42]
    \draw[myrectangle] (-15,5) rectangle (15,1.5);
    \node[mycircle=BrickRed] at (-14,4) {};
    \node[mycylinder] at (-8,4) {};
    \node[mycircle=OliveGreen] at (-2,4) {};
    \node[mycircle=OliveGreen] at (3,4) {};
    \node[mycylinder] at (8,4) {};
    \node[mycircle=BrickRed] at (14,4) {};
    \node[below] at (-8,3) {Classical takes two steps};
    \draw[myrectangle] (-15,-5) rectangle (0,0);
    \node[mycircle=OliveGreen] at (-14,-2) {};
    \fill[BrickRed!80,draw=none] (-14,-2) -- (-14,-1.4) arc (90:-90:0.6cm) -- cycle;
    \node[mycylinder] at (-8,-2) {};
    \node[mycircle=BrickRed] at (-2,-2) {};
    \fill[OliveGreen!80,draw=none] (-2,-2) -- (-2,-1.4) arc (90:-90:0.6cm) -- cycle;
    \node[below] at (-8,-3) {Quantum takes one step};
     \draw[myrectangle] (1,1) rectangle (15,-5);
    \node[mycylinder] at (6,0) {Inverter};
    \node[mycylinder] at (6,-2) {Constant 1};
    \node[mycylinder] at (6,-4) {Constant 2};
   
\end{tikzpicture}
    \caption{A Simple Experiment Demonstrates Superior Performance of Quantum Computing Over Classical Methods}
    \label{fig:simpleExample}
\end{figure}
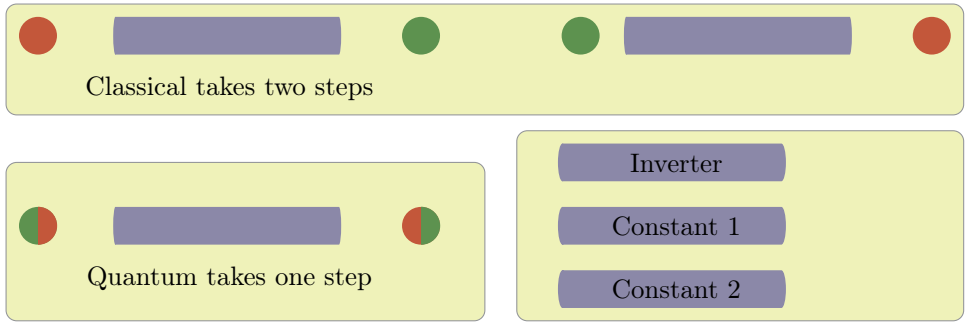

The superiority of quantum computing over classical computing can be explained through a simple illustration as shown in Figure \ref{fig:simpleExample}. First, we consider three identical-looking pipes. The first pipe functions as an inverter, transforming a red circle into green and vice versa. The second pipe, known as a constant 1 pipe, invariably changes any circle to green, irrespective of its initial colour. The third pipe, a constant 2, consistently changes any circle to red, regardless of its initial colour. The goal is to determine the type of randomly chosen pipe with the least number of attempts. We assume the pipe is an inverter, but this is not known beforehand. A classical computer requires at least two steps to identify the pipe. Initially, a red circle is passed through the pipe, which converts it to green. Subsequently, a green circle is sent through. If it appears as red, it verifies that the pipe is indeed an inverter. However, a quantum computer can accomplish the same task in just one step. A circle in a superposition of two colours, half red and half green, is passed through the pipe. The pipe converts the red half to green and the green half to red, resulting in a circle exhibiting both colours. This single step unequivocally reveals that the pipe is an inverter. Therefore, a quantum computer accomplishes the task in half the computational steps a classical computer requires. A similar technique is employed by the Deutsch algorithm \citep{deutsch1985quantum}. In more complex problems, quantum computers can provide exponential speedup \citep{qiu2018generalized}.

\subsection{Quantum gates and Circuits}

The fundamental building block of quantum information is the qubit. A qubit exhibits the potential to reside in state '0', state '1', and the superposition of these two fundamental states. In quantum mechanics, qubits are often denoted using the Dirac notation. The states '0' and '1' are depicted as vectors $| 0 \rangle $ and $ | 1 \rangle $, respectively, where $ | 0 \rangle = \begin{bmatrix}1 \\ 0\end{bmatrix} $ and $ | 1 \rangle = \begin{bmatrix}0 \\ 1\end{bmatrix} $.

In quantum computing, a qubit existing in a quantum state, often expressed as the superposition of states and denoted as $\Psi$, can be represented by the unitary vector:

\begin{equation}
|\Psi\rangle = \alpha| 0 \rangle + \beta| 1 \rangle \label{ch5:eq7}
\end{equation}

Here, $ \alpha $ and $ \beta $ are complex coefficients representing the probability amplitudes of the states $| 0 \rangle $ and $ | 1 \rangle $, respectively. When a qubit is measured, the likelihood of observing the '0' state is $ |\alpha|^2 $, and the probability of observing the '1' state is $ |\beta|^2 $, where $ \alpha, \beta \in C$ and $ |\alpha|^2 + |\beta|^2 = 1 $.

When multiple qubits are combined, they form a quantum register. A $ n $ qubits register can represent a superposition of $ 2^n $ classical states. Mathematically, this $ n $-qubit state can be expressed as:

\begin{equation}
|\Psi\rangle n = \sum{i=0}^{2^n-1} \alpha_i | X_i \rangle \label{ch5:eq8}
\end{equation}

Here, $ \alpha_i $ denotes the probability amplitude associated with the $ i^{th} $ state among all possible superposition states $ X_i $. It is worth noting that $ \sum_{i=0}^{2^n-1} |\alpha_i|^2 = 1 $.

In quantum computing, quantum gates perform the vital task of modifying qubit states.   Quantum gates are unitary matrices $U$ \citep{musz2013unitary} which act as operators to transform an input quantum state  $|\Psi_{in}\rangle$ to another state $|\Psi\rangle_{out}$. That is  $|\Psi_{out}\rangle=U|\Psi_{in}\rangle$. This transformation follows the following properties:
\begin{equation}
    UU^\dagger=U^\dagger U=I
\end{equation}
where $ I $ indicates identity matrix, and $ U^\dagger $ represents the complex conjugate of a matrix $ U $.

Unitary operations are characterized by two fundamental properties: reversibility and norm preservation \citep{williams2011quantum}. The property of reversibility implies that each unitary operation has a distinct inverse. On the other hand, preserving norms by a unitary operator signifies that the total probability remains one after the state transformation, thereby maintaining the superposition on the surface of the unit sphere.

\begin{figure}[!htb]
    \centering
    \begin{tikzpicture}[line cap=round, line join=round, >=Triangle]
  \clip(-2.19,-2.49) rectangle (2.66,2.58);
  \draw [shift={(0,0)}, lightgray, fill, fill opacity=0.1] (0,0) -- (56.7:0.4) arc (56.7:90.:0.4) -- cycle;
  \draw [shift={(0,0)}, lightgray, fill, fill opacity=0.1] (0,0) -- (-135.7:0.4) arc (-135.7:-33.2:0.4) -- cycle;
  \draw(0,0) circle (2cm);
  \draw [rotate around={0.:(0.,0.)},dash pattern=on 3pt off 3pt] (0,0) ellipse (2cm and 0.9cm);
  \draw (0,0)-- (0.70,1.07);
  \draw [->] (0,0) -- (0,2);
  \draw [->] (0,0) -- (-0.81,-0.79);
  \draw [->] (0,0) -- (2,0);
  \draw [dotted] (0.7,1)-- (0.7,-0.46);
  \draw [dotted] (0,0)-- (0.7,-0.46);
  \draw (-0.08,-0.3) node[anchor=north west] {$\varphi$};
  \draw (0.01,0.9) node[anchor=north west] {$\theta$};
  \draw (-1.01,-0.72) node[anchor=north west] {$\mathbf {\hat{x}}$};
  \draw (2.07,0.3) node[anchor=north west] {$\mathbf {\hat{y}}$};
  \draw (-0.5,2.6) node[anchor=north west] {$\mathbf {\hat{z}=|0\rangle}$};
  \draw (-0.4,-2) node[anchor=north west] {$-\mathbf {\hat{z}=|1\rangle}$};
  \draw (0.4,1.65) node[anchor=north west] {$|\psi\rangle$};
  \scriptsize
  \draw [fill] (0,0) circle (1.5pt);
  \draw [fill] (0.7,1.1) circle (0.5pt);
\end{tikzpicture}
\caption{The Bloch sphere: Any state of a qubit on the Bloch sphere can be represented as $|\Psi\rangle = \cos \frac{\theta}{2} |0\rangle + \exp^{i\phi} \sin \frac{\theta}{2} |1\rangle$, where $0 \leq \theta \leq \pi$, $0 \leq \phi \leq 2\pi$}
\label{fig:bloch}
\end{figure}
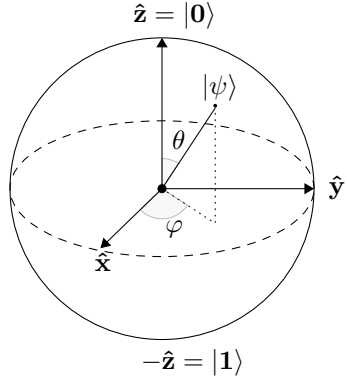

In quantum computing, qubit states are altered using gates. To comprehend the impact of these gates on qubit states, a useful geometric model known as the Bloch sphere \citep{mosseri2001geometry} is frequently employed. Figure \ref{fig:bloch} shows how quantum gates modify a qubit's state by rotating its position on the Bloch Sphere.

Different types of quantum gates are used to perform operations on qubits, which are the building blocks of quantum circuits \citep{Lambropoulos2007}. Quantum gates can be broadly classified into two categories: single-qubit gates and multi-qubit gates. Single-qubit gates operate on a single qubit at a time. Examples of these gates include the Pauli-X, Pauli-Y, and Pauli-Z gates, collectively known as Pauli gates, and the Hadamard gate. The Pauli gates are fundamental because their matrices are Hermitian, meaning that they are equal to their conjugate transpose and serve as the basis for more complex quantum operations. The Hadamard gate, on the other hand, is crucial for creating superposition states. Multi-qubit gates manipulate the states of two or more qubits simultaneously. A key example of a multi-qubit gate is the CNOT gate, which is vital in creating entanglement. Other important multi-qubit gates include the controlled-Z gate, the swap gate, and the Toffoli (CCNOT) gate. These gates are frequently used in constructing quantum circuits. Table \ref{table:gates} summarizes the primary quantum gates commonly utilized in constructing quantum circuits. Each gate's name, symbol, unitary matrix, and a brief description are presented.

\begin{longtable}{|p{1.5cm}|p{2.5cm}|p{3.5cm}|p{4cm}|}
\caption{Overview of Primary Quantum Gates } \label{table:gates} \\ 
\hline
\textbf{Gate Name} & \textbf{Symbol} & \textbf{Matrix} & \textbf{Description} \\ \hline
\endfirsthead

\multicolumn{4}{c}{{\tablename\ \thetable{} -- continued from previous page}} \\
\hline
\textbf{Gate Name} & \textbf{Symbol} & \textbf{Matrix} & \textbf{Description} \\ \hline
\endhead

\hline
\endfoot

\hline
\endlastfoot

Pauli $X$ gate &
  \begin{tikzcd} \qw &\gate{X} & \qw \end{tikzcd} & 
  \[
\begin{bmatrix}
    0 & 1 \\
    1 & 0 \\
\end{bmatrix}
\]     &
A single-qubit gate that inverts the state of a qubit. This gate performs a rotation of $\pi$ (180 degrees) about the x-axis\\ \hline

Rsotation gate ($R_X$) &
  \begin{tikzcd}
 \qw &\gate{R_X(\theta)} & \qw
\end{tikzcd} &
 \[
\begin{bmatrix}
   \cos\left(\frac{\theta}{2}\right) & -i\sin\left(\frac{\theta}{2}\right) \\ 
-i\sin\left(\frac{\theta}{2}\right) & \cos\left(\frac{\theta}{2}\right) 
\end{bmatrix}
\]   &
 Facilitates customized rotations around the x-axis of the Bloch sphere. \\ \hline

 Pauli $Y$ gate &
  \begin{tikzcd}\qw &\gate{Y} & \qw\end{tikzcd} &
  \[
\begin{bmatrix}
    0 & -i \\
    i & 0 \\
\end{bmatrix}
\] &
  A rotation of $\pi$ (180 degrees) about the y-axis of the Bloch sphere \\ \hline

$Y$ rotation gate ($R_Y$) &
  \begin{tikzcd}
 \qw &\gate{R_Y(\theta)} & \qw
\end{tikzcd} &
  \[\begin{bmatrix}
  \cos\left(\frac{\theta}{2}\right) & -\sin\left(\frac{\theta}{2}\right) \\ 
\sin\left(\frac{\theta}{2}\right) & \cos\left(\frac{\theta}{2}\right) 
\end{bmatrix}
\]   &
  Customized rotations around the Y-axis of the Bloch sphere.\\ \hline

Pauli $Z$ gate  &
  \begin{tikzcd}\qw &\gate{Z} & \qw\end{tikzcd} &
  \[
\begin{bmatrix}
    1 & 0 \\
    0 & -1 \\
\end{bmatrix}
\] &
 A rotation of $\pi$ (equivalent to 180 degrees) on the input state around the z-axis of the Bloch sphere. It is sometimes referred to as the sign flip gate.  \\ \hline

 $R_z$ gate &
 \begin{tikzcd}
 \qw &\gate{R_Z(\theta)} & \qw
\end{tikzcd}  &
  \[\begin{bmatrix}
   e^{-i\frac{\theta}{2}} & 0 \\ 
0 & e^{i\frac{\theta}{2}} 
\end{bmatrix}
\] &
 A rotation of $\pi$ (equivalent to 180 degrees) on the input state around the z-axis of the Bloch sphere.  \\ \hline

S gate &
  \begin{tikzcd}\qw &\gate{S} & \qw\end{tikzcd} &
  \[
\begin{bmatrix}
1 & 0 \\ 
0 & i  
\end{bmatrix}
\]   &
  A phase gate that performs a rotation of $\pi/2$ (90 degrees) on the input state around the z-axis of the Bloch sphere. The $S^\dagger$ gate, also known as the conjugate transpose or the adjoint of the $S$ gate, performs the inverse operation, effectively negating the action of the $S$ gate. \\ \hline

T gate &
  \begin{tikzcd}\qw &\gate{T} & \qw\end{tikzcd} &
  \[
\begin{bmatrix}
  1 & 0 \\ 
0 & e^{i\pi/4} 
\end{bmatrix}
\]   &
  A rotation of $4\pi/4$ (45 degrees) on the input state around the z-axis of the Bloch sphere. The $T^\dagger$ gate, which is the conjugate transpose or the adjoint of the $T$ gate, performs the inverse operation, effectively negating the action of the $T$ gate. \\ \hline

Hadamard gate &
  \begin{tikzcd}\qw &\gate{H} & \qw\end{tikzcd} &
  \[
\frac{1}{\sqrt{2}} \begin{bmatrix}
    1 & 1 \\
     1 & -1 \\
\end{bmatrix}
\]  &
  A fundamental quantum gate that is both Hermitian and unitary. It is particularly known for its ability to create superposition, which is a key aspect of quantum computing.\\ \hline
  
Measure-ment gate &
  \begin{tikzcd}\qw & \meter{} & \qw\end{tikzcd} &
  -- &
 Measurement gates are used to obtain classical data from a quantum system.  When a quantum system is measured, it collapses from a superposition state into one of its possible classical states, with the probabilities of each outcome determined by the quantum state itself. \\ \hline

CNOT gate &
\begin{tikzcd}
& \ctrl{1} & \qw \\
 & \targ{} & \qw
\end{tikzcd}
& \[
\begin{bmatrix}
1 & 0 & 0 & 0 \\
0 & 1 & 0 & 0 \\
0 & 0 & 0 & 1 \\
0 & 0 & 1 & 0 \\
\end{bmatrix}
\] &
   The Controlled-NOT gate, also known as the quantum CX gate, is a two-qubit gate that plays a crucial role in quantum computing. It alters the state of the target qubit if the control qubit is in the $|1\rangle$ state, and leaves it as is if the control qubit is in the $|0\rangle$ state.\\ \hline

CZ gate &
  \begin{tikzcd}
& \ctrl{1} & \qw \\
 &  \control{} & \qw
\end{tikzcd}
&  \[\begin{bmatrix}
1 & 0 & 0 & 0 \\
0 & 1 & 0 & 0 \\
0 & 0 & 1 & 0 \\
0 & 0 & 0 & -1 \\
\end{bmatrix}
\] &
 The quantum Controlled-Z (CZ) gate operates much like the CNOT gate and is alternatively known as the CPHASE gate. It changes the phase of the target qubit (a phase flip) only if the control qubit is in the state $|1\rangle$; otherwise, it leaves the target qubit unchanged. \\ \hline

CCNOT gate &
\begin{tikzcd}
& \ctrl{2} & \qw \\
& \ctrl{1} & \qw \\
 & \targ{} & \qw
\end{tikzcd}
&  \[
\begin{bmatrix}
1 & 0 & 0 & 0 & 0 & 0 & 0 & 0 \\
0 & 1 & 0 & 0 & 0 & 0 & 0 & 0 \\
0 & 0 & 1 & 0 & 0 & 0 & 0 & 0 \\
0 & 0 & 0 & 1 & 0 & 0 & 0 & 0 \\
0 & 0 & 0 & 0 & 1 & 0 & 0 & 0 \\
0 & 0 & 0 & 0 & 0 & 1 & 0 & 0 \\
0 & 0 & 0 & 0 & 0 & 0 & 0 & 1 \\
0 & 0 & 0 & 0 & 0 & 0 & 1 & 0 \\
\end{bmatrix}
\]
 &
 The Toffoli gate (also known as CCX or CCNOT) is a three-qubit quantum gate that extends the CNOT gate. Inverts the state of the third qubit only if the first two qubits are in the $|1\rangle$ state. \\ \hline

 SWAP gate &
\begin{tikzcd}
&  \swap{1} & \qw \\
& \targX{}  & \qw \\
\end{tikzcd}
& \[ 
\begin{bmatrix}
1 & 0 & 0 & 0 \\
0 & 0 & 1 & 0 \\
0 & 1 & 0 & 0 \\
0 & 0 & 0 & 1 \\
\end{bmatrix}
\] &
 The SWAP gate is used in quantum computing to exchange the states of two qubits. For example, the state $|01\rangle$ would become $|10\rangle$, and vice versa. \\ \hline

CSWAP gate &
\begin{tikzcd}
&  \ctrl{1} & \qw \\
&  \swap{1} & \qw \\
& \targX{}  & \qw \\
\end{tikzcd}
&  \[
\begin{bmatrix}
1 & 0 & 0 & 0 & 0 & 0 & 0 & 0 \\
0 & 1 & 0 & 0 & 0 & 0 & 0 & 0 \\
0 & 0 & 1 & 0 & 0 & 0 & 0 & 0 \\
0 & 0 & 0 & 1 & 0 & 0 & 0 & 0 \\
0 & 0 & 0 & 0 & 1 & 0 & 0 & 0 \\
0 & 0 & 0 & 0 & 0 & 0 & 1 & 0 \\
0 & 0 & 0 & 0 & 0 & 1 & 0 & 0 \\
0 & 0 & 0 & 0 & 0 & 0 & 0 & 1 \\
\end{bmatrix}
\] &
 The Controlled-SWAP (CSWAP) gate, also known as the Fredkin gate,  is a controlled version of the SWAP gate. It operates on three qubits. When the first qubit (the control qubit) is in the state $|1\rangle$, the states of the other two qubits are swapped. \\ \hline

\end{longtable}
 
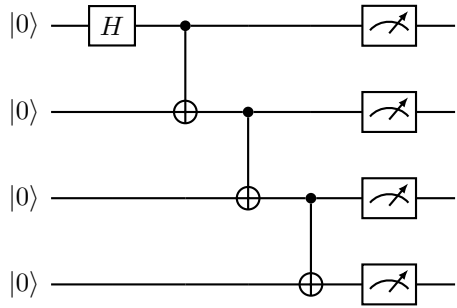
\begin{figure}[!htb]
\centering
\begin{tikzcd}
\lstick{\ket{0}} & \gate{H}  & \ctrl{1} & \qw &\qw & \meter{} & \qw \\
\lstick{\ket{0}} & \qw & \targ{} & \ctrl{1} & \qw & \meter{} & \qw \\
\lstick{\ket{0}} & \qw & \qw & \targ{} & \ctrl{1} & \meter{} & \qw \\
\lstick{\ket{0}} &  \qw  & \qw & \qw & \targ{} & \meter{} & \qw \\
\end{tikzcd}
\caption{A 4-qubit GHZ state Circuit}
	\label{Q_Circuit}
\end{figure}

 A quantum circuit is formed when a sequence of quantum gates is applied to one or multiple qubits. These circuits manipulate and process quantum information to achieve specific computational goals. Typically, a quantum circuit undergoes five primary stages, as outlined in \citep{Roman19}. Qubits are prepared in a defined initial state in the beginning stage, often the $|0\rangle$ state. This is followed by the Superposition stage, where quantum gates like the Hadamard gates are used to create a superposition of multiple possible states. The third stage is the quantum algorithm implementation, where a sequence of single- and multi-qubit gates perform the core computation, entangling qubits and evolving the state. Then, interference increases the likelihood of identifying the correct solution by manipulating quantum interference in the qubit state. Finally, the measurement gate collapses the superposition, yielding a classical outcome representing the result of the quantum computation. Figure \ref{Q_Circuit} depicts a 4-qubit quantum circuit designed to generate a Greenberger–Horne–Zeilinger (GHZ) state. The circuit begins with all qubits in the $|0\rangle$ state and, through the application of specific quantum gates, produces the entangled GHZ state $|GHZ\rangle = \frac{1}{\sqrt{2}}(|0000\rangle + |1111\rangle)$.

\subsection{Quantum Programming and Quantum SDKs}

Quantum programming languages are essential for translating concepts into executable instructions for quantum computers \citep{heim2020quantum}. Leading quantum computing vendors have launched their quantum SDKs to help developers, and researchers harness the power of quantum computers easily. These SDKs provide tools for creating, manipulating, and simulating quantum programs. Like in classical computing, quantum programming utilizes high- and low-level languages.
High-level languages such as Q\#, Qiskit, and Cirq are designed to integrate with widely used classical programming environments, allowing developers to quickly write and execute code and combine quantum algorithms with classical computing tasks.  \citep{serrano2022quantum}. Conversely, low-level languages, like cQASM, Quil, and OpenQASM, transform complex algorithms from a higher level into tangible operations that are carried out on quantum processors. Figure \ref{fig:code-snippet} shows a code snippet of an entangled state using Qiskit language. The code provided utilizes the QASM simulator from the IBM Aer library to create an entangled state. The quantum circuit is executed multiple times, generating either the state `00' or `11' with each run. The code output shows that out of 1000 runs, `11' was measured 490 times, and `00' was measured 510 times. Table \ref{table:sdks} summarizes some well-known quantum  SDKs that major quantum computing companies utilize. For more details, review papers \citep{fingerhuth2018open} and \citep{serrano2022quantum} have highlighted open-source software projects in quantum computing.

\begin{figure}[!htb]
    \centering
    \lstdefinestyle{mystyle}{
    language=Python,
    basicstyle=\ttfamily\small,
    keywordstyle=\color{blue},
    commentstyle=\color{green},
    stringstyle=\color{red},
    numbers=left,
    numberstyle=\tiny\color{gray},
    breaklines=true,
    breakatwhitespace=true,
    showspaces=false,
    showstringspaces=false
}
\begin{lstlisting}[style=mystyle]
from qiskit import QuantumCircuit, Aer, execute
# Create a two-qubit circuit
circuit = QuantumCircuit(2, 2)
# Apply a Hadamard gate to the first qubit
circuit.h(0)
# Apply a CNOT gate between the first and second qubits
circuit.cx(0, 1)
# Measure the qubits
circuit.measure([0, 1], [0, 1])
# Simulate the circuit
simulator = Aer.get_backend('qasm_simulator')
job = execute(circuit, simulator, shots=1000)
result = job.result()
# Get the counts
counts = result.get_counts(circuit)
print("Measurement results:", counts)
\end{lstlisting}
    \caption{Code snippet of an entangled state using Qiskit}
    \label{fig:code-snippet}
\end{figure}

\begin{longtable}{|p{1.5cm}|p{2.5cm}|p{3.5cm}|p{4cm}|}
\caption{Exploring SDKs in Prominent Quantum Computing Platforms} \label{table:sdks} \\ 
\hline
\textbf{Name} & \textbf{Vendor} & \textbf{Implemented Programming Language} & \textbf{Description} \\ \hline
\endfirsthead

\multicolumn{4}{c}{{\tablename\ \thetable{} -- continued from previous page}} \\
\hline
\textbf{Name} & \textbf{Vendor} & \textbf{Implemented Programming Language} & \textbf{Description} \\ \hline
\endhead

\hline
\endfoot

\hline
\endlastfoot

Ocean &
  D-Wave &
  Python &
An open-source SDK specifically designed for D-Wave's quantum computers, with a focus on quantum annealing systems. This SDK is well-suited for solving optimization and sampling problems. \\ \hline
Qiskit &
  IBM &
  Python &
 An open-source SDK designed specifically for universal gate-based quantum computing provides a suite of tools for circuit design, simulation, and deployment on IBM quantum hardware.\\ \hline
Cirq &
  Google &
  Python &
  An open-source framework designed for creating and manipulating NISQ circuits. \\ \hline
Forest (pyQuil) &
  Rigetti &
  Python &
  An SDK specifically designed for Rigetti’s gate-based quantum computers, featuring a robust Quil instruction language.\\ \hline
Braket SDK &
  Amazon &
  Python &
 An SDK that provides access to Amazon’s quantum computing services, including simulators, managed devices, and hybrid quantum-classical algorithms.  \\  \hline
Strawberry Fields/Blackbird &
  Xanadu &
  Python &
Libraries specifically designed for continuous variable photonic quantum computing.  \\ \hline
Quantum Development Kit (QDK) &
  Microsoft &
  Python, Q\# &
 A set of tools designed to build quantum applications. It incorporates Q\#  along with utilities for simulating and estimating resources. \\ \hline
Orquestra &
  Zapata &
  Python &
  This is a platform that concentrates on the development and deployment of quantum-classical workflows in the near term, with a particular focus on applications for businesses. \\ \hline
IonQ &
  IonQ &
  Python &
  An native SDK for IonQ’s trapped-ion quantum computers, providing access to both gate-based and native-gate models.\\ \hline
ProjectQ &
  ETH Zurich &
  Python &
 An open-source structure that is intended for the modular creation and compilation of quantum algorithms. \\ \hline
PennyLane &
  Xanadu &
  Python &
   An SDK  is designed for QML, differentiable quantum programming, and hybrid quantum-classical models. It is a cross-platform library, making it versatile for various quantum computing applications\\ \hline
Azure Quantum &
  Microsoft &
  Python, Q\# &
 This is a cloud-based environment designed for quantum solutions, offering access to a variety of hardware providers and tools for quantum development. \\ \hline

\end{longtable}


\subsection{Quantum Algorithms}
Quantum algorithms are step-by-step procedures to solve a particular problem efficiently on a quantum machine. Most of the quantum algorithms are designed considering the execution of a universal quantum gate model. We have highlighted the major quantum algorithms and their probable applications.

Grover’s algorithm, a quantum search algorithm designed by Lov Grover in 1996, can efficiently search an unsorted database \citep{grover1996fast}. If there are $N$ items in the unsorted database, this algorithm can find a specific item using $O(\sqrt{N})$ queries. Providing a quadratic speedup over classical algorithms, Grover’s algorithm is used as a powerful tool in various areas, including cryptography, optimization problems, and machine learning.

The Deutsch algorithm, introduced by David Deutsch in 1985 \citep{deutsch1985quantum}, served as an essential foundation for quantum algorithms. Its generalized form, the Deutsch-Jozsa algorithm, was later developed by David Deutsch and Richard Jozsa in 1992 \citep{deutsch1992rapid}. The Deutsch-Jozsa algorithm distinguishes between constant and balanced functions through a single-function evaluation. In this context, a continuous function yields outputs of 0 or all 1, while a balanced function exhibits an equal number of 1s and 0s in the outputs. The Deutsch-Jozsa algorithm exemplifies significantly faster computation than classical algorithms when addressing constant-balanced problems.

Another robust quantum algorithm is Shor’s algorithm \citep{shor1999polynomial}, developed by Peter Shor in 1994. It has the remarkable ability to identify the prime factors of an integer using a quantum computer in polynomial time, which is exponentially faster than the most efficient known classical algorithms. The implications of Shor’s algorithm are profound, particularly in its potential to compromise RSA encryption and similar cryptographic systems.

The Harrow-Hassidim-Lloyd (HHL) algorithm, developed by Aram W. Harrow, Avinatan Hassidim, and Seth Lloyd in 2009, is utilized for solving systems of linear equations \citep{harrow2009quantum}. It is particularly effective with large, sparse matrices and can outperform classical algorithms under certain conditions. This algorithm has potential applications in various fields, including QML, optimization, and quantum simulations.

In addition to the algorithms mentioned above, numerous other quantum computing algorithms have been developed. We have discussed some of these algorithms in the following sections. The development of many more algorithms is underway. However, implementing quantum algorithms on current NISQ devices is challenging due to their limited resources, qubit decoherence, and the need for substantial classical preprocessing  \citep{leymann2020bitter}.


\section{Methodology of Searching Articles}

This section presents our selection process for articles based on a systematic review of the literature that intersects quantum computing and software engineering. We used a systematic literature review to ensure a fair, credible, and unbiased evaluation of methodologies for retrieving information related to QSE and quantum-driven software engineering. It should be noted that our paper provides an overview of quantum computing, prominent quantum algorithms, and quantum SDKs, but we did not incorporate a systematic literature review for accessing that information. During the systematic review of the literature, we classified our search into three broad groups and ensured that the selected papers for each group contained relevant information:

\begin{itemize}
\item Group 1: The application of QML in software engineering tasks.

 \item Group 2: The utilization of quantum optimization and quantum algorithms in software engineering tasks.

 \item Group 3: Tools and techniques of QSE practices.
\end{itemize}

For searching for relevant articles, we use six prominent academic research databases including IEEE Xplore \footnote{https://ieeexplore.ieee.org/Xplore/home.jsp}, Science Direct \footnote{https://www.sciencedirect.com/},  Springer Link \footnote{https://link.springer.com/}, ACM Digital Library \footnote{https://dl.acm.org/}, Scopus \footnote{https://www.scopus.com/search/form.uri?display=basic\#basic} and Web of Science \footnote{https://www.webofscience.com/wos/woscc/basic-search}. These databases are considered because they present scientifically and technically peer-reviewed, high-quality papers in their digital libraries. The search process includes journal articles, conference and workshop proceedings, and book chapters. In addition to searching within these databases, we also look for relevant articles from key conferences and workshops related to quantum computing and software engineering. This helps us find any articles that might have been missed during the database search. The conferences and workshops that we focus on are as follows:

\begin{itemize}
    \item IEEE International Conference on Quantum Software (QSW)
    \item IEEE/ACM International Workshop on Quantum Software Engineering (QSE)
    \item IEEE International Conference on Quantum Computing and Engineering (QCE)
    \item International Conference on Software Engineering Workshops (ICSEW) 
    \item International Workshop on Architectures and Paradigms for Engineering Quantum Software
    
\end{itemize}

Given our interest in the relatively latest research articles, we limited our search to those published in the past twenty years, from 2004 to 2023. To manage this extensive collection, we rely on Mendeley \footnote {https://www.mendeley.com/}  as a bibliographic manager. Mendeley helps us organize, read, and annotate articles, as well as remove duplicates and filter out irrelevant entries.

The query strings used for the search for three groups are as follows:

\begin{itemize}
\item Group 1: 
("Quantum machine learning" OR "Quantum Neural Network" OR "Quantum Support Vector Machines" OR "Variational Quantum" OR "Quantum Principal Component Analysis") AND ("Software Engineering" OR "Software Bug prediction" OR "Software Defect Prediction" OR "Software Reliability Prediction" OR "Vulnerability Prediction" OR "Software Quality Prediction" OR "Software Requirements Prediction" OR "Software Usability Prediction" OR "Code Smell")

 \item Group 2: ("Quantum optimization" OR "Quantum annealing" OR "quantum search" OR "Variational Quantum" OR "QUBO" OR "D-wave" OR "QAOA" OR "VQE" OR "QUDO") AND ("Software
Engineering" OR "Software testing" OR "Software verification" OR "static analysis" OR "Software model
checking" OR "false path pruning" OR "test suite reduction" OR "code clone ")

 \item Group 3: ("Quantum Software Engineering" OR "Quantum Software Development" OR "quantum search" OR "Quantum Testing and
debugging" OR "Quantum software
modeling" OR "Quantum Software Engineering process" OR "Quantum Testing and
debugging" OR "Quantum Software Engineering Tools" OR "Refactoring" OR "Reverse engineering", "Quantum UML") 
\end{itemize}

Using the search keys, we individually search each dataset for different groups. The same article can commonly be present in multiple datasets. We identify and eliminate duplicate articles retrieved from the databases. Then, with manual search in conferences and workshops related to the domain, we include articles if they were missed in the query-based search from the database. 

The following inclusion and exclusion criteria are defined:
\begin{itemize} 
\item IC1: The selected papers must deal with both software engineering and quantum computing.
\item IC2: The articles must be published in a journal, conference or workshop proceedings, or book chapters. 
\end{itemize} 
Exclusion criteria: 
\begin{itemize} 
\item EC1: The study focuses on either software engineering or quantum computing, not both.
\item EC2: The papers that are not written in English.
\item EC3: The study that was published before 2004.
\end{itemize}

\begin{figure}[!htb]
	\centering
	\pgfplotsset{compat=1.11,
		/pgfplots/ybar legend/.style={
			/pgfplots/legend image code/.code={%
				\draw[##1,/tikz/.cd,yshift=-0.25em]
				(0cm,0cm) rectangle (3pt,0.8em);},
		},
	}
	\pgfplotstableread[col sep=comma]{data.csv}\datatable
	\begin{tikzpicture}
	\begin{axis}[width=15cm,height=6cm,
	enlargelimits=0.03,
	ybar=0.1*\pgflinewidth,
	enlarge x limits=0.02,
	ymin=0,
	ymax=14,
	ytick={0,1,2,3,4,5,6,7,8,9,10,11,12,13,14,15},
	bar width=3.5pt,
	xtick=data,
	x tick label style={rotate=90,anchor=east},
	xticklabels from table={\datatable}{Year},
	axis background/.style={fill=white},
	ymajorgrids=true,
	major x tick style=transparent,
	legend style={legend columns=3, at={(0.7, 1.0)}},
	xlabel={Year of Publication},
	ylabel={Number of Publications},
	]
	\addplot [fill=red] table [x expr=\coordindex, y=Group 1]{\datatable};
	\addplot [fill=black] table [x expr=\coordindex, y=Group 2]{\datatable};
	\addplot [fill=green] table [x expr=\coordindex, y=Group 3]{\datatable};
	\legend{Group 1,Group 2,Group 3}
	\end{axis}
	\end{tikzpicture}
	\caption{Publication Counts Over the Years for Three Groups (Group 1, Group 2, Group 3)}
	\label{fig:pubdistribution}
\end{figure}
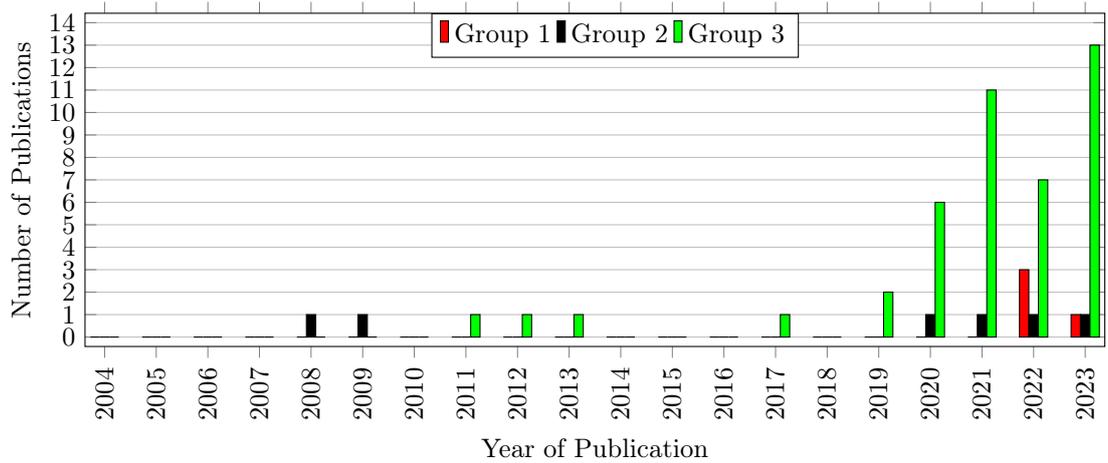

After selecting, we identified approximately 12 articles from Group 1, 17 from Group 2, and 98 from Group 3 that satisfied our set criteria. We manually checked the papers for quality and relevance, especially when we read the abstract, methodology, and findings, checking the relevance associated with the group requirements. When a paper appears in multiple formats, such as a journal article and a conference proceeding or book chapter, we select the version that provides the most comprehensive details. As a result, we ended up with 56 publications, with approximately 4 articles for Group 1, 6 for Group 2, and 43 for Group 3. Figure \ref{fig:pubdistribution} illustrates the distribution of publications in different years for three groups.

\section{Quantum computing for Software Engineering Research}

In this section, we explore how quantum computing can be utilized in software engineering. We examine the potential of QML techniques to expedite various software engineering tasks and the use of quantum search and optimization algorithms to address challenges within the field. Our main goal is to highlight the state-of-the-art and potential advancements quantum computing offers for software engineering research.

\subsection{QML for Software Engineering Research}

\subsubsection{Background of QML}

\begin{figure}
    \centering
    \begin{tikzpicture}
		\definecolor{lightblue}{RGB}{169,214,235}
		\definecolor{lightpurple}{RGB}{221,189,246}
		\definecolor{lightpink}{RGB}{249,192,192}
		\definecolor{lightturquoise}{RGB}{186,232,232}
		
		\fill[lightblue] (0,2) rectangle (2,4);
		\fill[lightpurple] (2,2) rectangle (4,4);
		\fill[lightpink] (0,0) rectangle (2,2);
		\fill[lightturquoise] (2,0) rectangle (4,2);
		
		\draw (0,0) rectangle (4,4);
		\draw (2,0) -- (2,4);
		\draw (0,2) -- (4,2);
		
		\node at (1,3) {\Large CC};
		\node at (3,3) {\Large CQ};
		\node at (1,1) {\Large QC};
		\node at (3,1) {\Large QQ};
		
		\draw[->] (0,4.5) -- (4,4.5); 
		\node[above] at (2,4.5) {Type of Algorithm};
		\draw[->] (-0.8,0) -- (-0.8,4); 
		\node[rotate=90] at (-1.3,2) {Type of Data};
		
		\node[rotate=90, left] at (-0.5,4) {Classical};
		\node[rotate=90, left] at (-0.5,2) {Quantum};
		\node[above] at (1,4) {Classical};
		\node[above] at (3,4) {Quantum};
	\end{tikzpicture}
    \caption{Four unique methods for integrating quantum computing with machine learning. \citep{Chen31122024}}
    \label{fig:qmltaxonomy}
\end{figure}
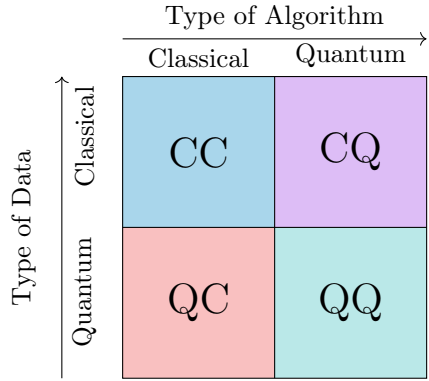

Machine learning is a field of computer science that encompasses algorithms designed to analyze data and detect patterns within it. QML is a burgeoning discipline that fuses the principles of quantum computing with traditional machine learning \citep{schuld2015introduction}. The advantages of QML include the capacity to manage large data sets with greater efficiency and deliver faster solutions. It also improves the speed and accuracy of data-driven learning. Based on the nature of the data and processing device used, QML is broadly classified into four categories \citep{aimeur2006machine}. In the Classical-Classical (CC) category, classical data undergoes processing using conventional computing methods but integrates concepts inspired by quantum mechanics.
On the other hand, in the Quantum-Classical (QC) category, quantum computing is employed to handle classical data. The Classical-Quantum (CQ) category encompasses machine learning that relies on classical data but integrates quantum computing either within the model or during training. In contrast, the Quantum-Quantum (QQ) type involves QML, where both input data and processing methods are inherently quantum. Among these four types, the CQ and CC methods have received a more extensive exploration in the field of QML \citep{sood2023quantum}.  Figure \ref{fig:qmltaxonomy} represents four unique methods for integrating quantum computing with machine learning.

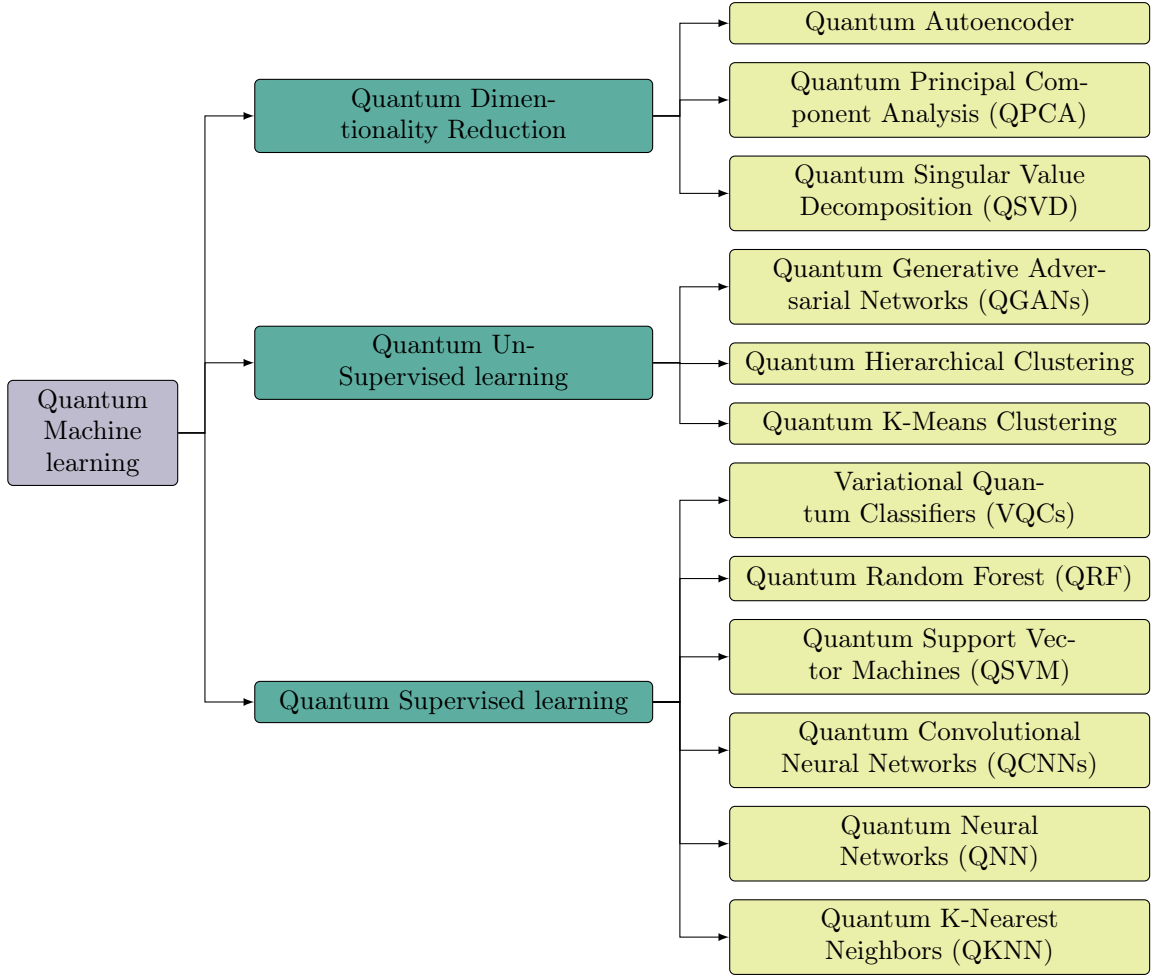
\begin{figure}[!htb]
    \centering 
\tikzset{
    basic/.style  = {draw, text width=2cm, align=center, rectangle,fill=CadetBlue!40,rounded corners=2pt,},
    tnode/.style = {basic, thin, align=left, fill=GreenYellow!60, text width=15em, align=center,rounded corners=2pt,},
    xnode/.style = {basic, thin, rounded corners=2pt, align=center, fill=PineGreen!60,text width=5cm,},
}
\begin{forest} for tree={
    grow=east,
    growth parent anchor=west,
    parent anchor=east,
    child anchor=west,
    edge path={\noexpand\path[\forestoption{edge},->, >={latex}] 
         (!u.parent anchor) -- +(10pt,0pt) |-  (.child anchor) 
         \forestoption{edge label};}
}
[Quantum Machine learning, basic,  l sep=10mm,
    [Quantum Supervised learning, xnode,  l sep=10mm,
        [Quantum K-Nearest Neighbors (QKNN), tnode]
        [Quantum Neural Networks (QNN), tnode]
        [Quantum Convolutional Neural Networks (QCNNs), tnode]
        [Quantum Support Vector Machines (QSVM), tnode]
        [Quantum Random Forest (QRF), tnode]
        [Variational Quantum Classifiers (VQCs), tnode]
        ]
    [Quantum Un-Supervised learning, xnode,  l sep=10mm,
        [Quantum K-Means Clustering, tnode]
        [Quantum Hierarchical Clustering, tnode] 
        [Quantum Generative Adversarial Networks (QGANs), tnode] 
        ]
    [Quantum Dimentionality Reduction, xnode,  l sep=10mm,
        [Quantum Singular Value Decomposition (QSVD), tnode]
        [Quantum Principal Component Analysis (QPCA), tnode]
        [Quantum Autoencoder, tnode] 
         ] 
    ]
\end{forest}
    \caption{Different Quantum Machine learning approaches}
    \label{fig:qml_type}
\end{figure}

QML has several strategies: fully quantum machine learning, hybrid classical-quantum machine learning, quantum-enhanced machine learning, and quantum-inspired machine learning \citep{HOUSSEIN2022116512}. Fully quantum machine learning involves algorithms that are entirely run on quantum computers, which are largely theoretical at this point. Hybrid classical-quantum machine learning employs variational quantum circuits where a quantum circuit, parameterized, learns to represent data or a model, and classical optimization is used to adjust the circuit parameters for iterative performance improvement. Quantum-enhanced machine learning aims to use quantum resources to enhance classical machine learning tasks like optimization, feature selection, and data preprocessing. An example would be using quantum linear algebra routines to boost classical machine learning algorithms. Finally, quantum-inspired machine learning algorithms apply principles of quantum mechanics to enhance traditional machine learning algorithms while still running on classical computers. Considering the abovementioned strategies, several quantum variants of classical machine learning algorithms have been proposed in the literature. Given these strategies, a single classical machine learning algorithm can have multiple quantum counterparts. For instance, neural networks have four different variants. \citep{jeswal2019recent}.  Figure \ref{fig:qml_type} summarizes commonly utilized QML approaches in the literature. These approaches include supervised, unsupervised, dimensionality reduction methods. It is noted that major quantum software development providers such as PennyLane (Xanadu), TensorFlow Quantum (Google), Qiskit (IBM), Cirq (Google), and Ocean SDK (D-Wave) have developed several powerful tools and frameworks for implementing QML.

\begin{figure}[!htb]
    \centering 
    \begin{tikzpicture}
        \node[draw, fill=green, rounded corners, fill=Emerald!70] (quantum) {
            \begin{quantikz}
                \lstick{$\ket{q_0}$}  & \gate[5]{\mu_\phi(x) \qquad}   & \gate[5]{U(\theta)}  & \meter{} \\
                \lstick{$\ket{q_1}$}  &                                 &                      & \meter{} \\
                \lstick{$\ket{q_2}$}  &                                 &                      & \meter{} \\
                \vdots                 &                                 &                      & \vdots \\
                \lstick{$\ket{q_n}$}  & \gateinput{Feature map}       & \gateinput{Ansatz}   & \meter{}
            \end{quantikz}
        };
        
        \node [draw, minimum width=3.5cm, minimum height=6cm, right=of quantum, 
               rounded corners, fill=GreenYellow!70] (main) {};

        \node [draw, below left of=main, xshift=0.8cm, yshift=2.3cm, 
               minimum width=2.5cm, minimum height=1.5cm, rectangle, fill=Cyan!70, 
               text width=5em, text centered, rounded corners] (sub1) 
               {Evaluate Cost Function $f(\theta)$};
        
        \node [draw, below left of=main, xshift=0.8cm, yshift=-1cm, 
               minimum width=2.5cm, minimum height=1.5cm, rectangle, fill=Cyan!50, 
               text width=5em, text centered, rounded corners] (sub2) 
               {Classical Optimization};

        \draw [->] (quantum) -- (main);
        \draw [->] (sub1) -- (sub2);
        \draw [->] (main.south) -- +(0cm,-1cm) -| ([xshift=0.85cm,yshift=-2.8cm]quantum.south) 
              node[midway, above, xshift=3cm] {update ($\theta$)};
        
        \node[above=of quantum, yshift=-2.5em] {Quantum part};
        \node[above=of main, yshift=-2.5em] {Classical part};

    \end{tikzpicture}
    \caption{Basic structure of Variational Quantum Classifier}
    \label{fig:vqc}
\end{figure}
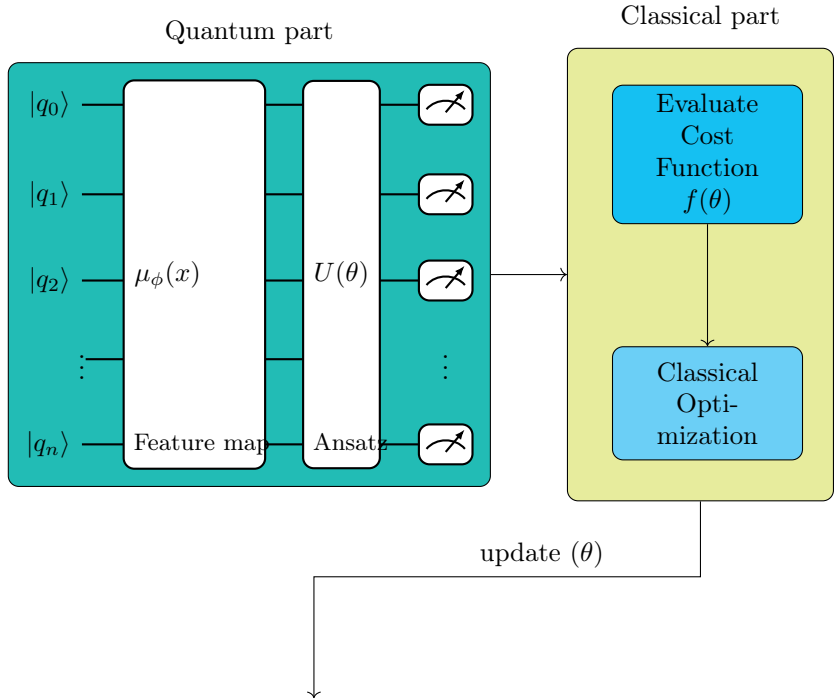

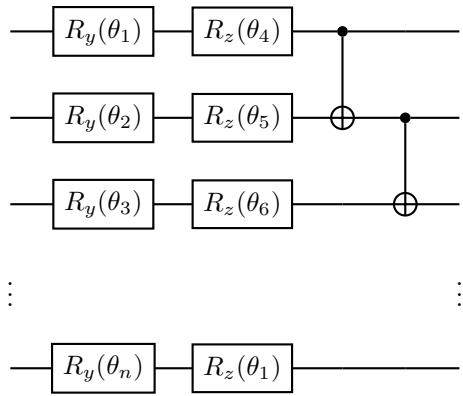
\begin{figure}[!htb]
    \centering 
    \begin{quantikz}
        & \gate{R_y(\theta_1)}  & \gate{R_z(\theta_4)} &\ctrl{1} & \qw & \qw \\
        & \gate{R_y(\theta_2)} & \gate{R_z(\theta_5)} & \targ{} & \ctrl{1} & \qw \\
        & \gate{R_y(\theta_3)} & \gate{R_z(\theta_6)} & \qw & \targ{} & \qw \\
        \vdots & & & & & \vdots \\
        & \gate{R_y(\theta_n)} & \gate{R_z(\theta_1)} & \qw & \qw & \qw
    \end{quantikz}
    \caption{An example of an ansatz}
    \label{fig:ansatz}
\end{figure}

Among the QML approaches, the hybrid quantum-classical method is particularly suitable for current NISQ devices \citep{callison2022hybrid}. This is because the classical component efficiently utilizes the limited qubits available in NISQ devices and helps mitigate their noise. One of the fundamental hybrid quantum-classical QML algorithms for classification problems is the Variational Quantum Classifier (VQC) \citep{Miyahara2022}. Here, the circuits typically consist of layers of quantum gates, with the parameters of these gates adjusted iteratively through a classical optimization algorithm, such as gradient descent, to enhance the classifier’s accuracy. In VQC, classical data are initially encoded as quantum states, which can be achieved with feature mapping. The next step involves constructing a variational quantum circuit composed of parameterized quantum gates that manipulate the quantum states, also known as an ansatz \citep{Cerezo2021}. This circuit introduces some degree of superposition and entanglement. Subsequently, the qubits undergo measurement, and the measurement outcomes are post-processed and input into a classical cost function. Through an iterative process, the learnable parameters of the unitary transformation are refined until the cost function is minimized. Figure \ref{fig:vqc} illustrates how a VQC is used for classification tasks. In contrast, Figure \ref{fig:ansatz} shows a simple ansatz constructed using two different rotation gates, $R_y$ and $R_z$, with entanglement.

\subsubsection{Application of QML in Software Engineering research}

In recent years, QML has gained significant popularity due to its potential to address numerous real-world problems. These include natural language processing, recommendation systems, speech recognition, image classification, and applications within the medical domain. Additionally, efforts have been made to explore the applicability of QML to addressing challenges within the realm of software engineering.

In their seminal study,  \cite{Zhou2022} presented the potential of Quantum Neural Networks (QNNs) in addressing software vulnerability detection challenges. Usually, the increased capacity, higher effective dimensions, faster training, and superior learning ability of QNNs make them better suited for large-scale prediction tasks than their classical counterparts. Based on this motivation, they proposed a Quantum Deep Embedding Neural Network (QDENN) for vulnerability detection, which effectively addresses batch processing challenges with long sentences and inconsistent input length inherent in QNNs. The QDENN comprises three components: a series of quantum gates with non-adaptable parameters for state preparation, a series of quantum gates with adaptable parameters to mimic biological neurons, and a set of measurement operations to extract output. Compared to other QNNs, the proposed QDENN achieves a higher vulnerability detection accuracy, with the best average vulnerability detection accuracy of the model reaching 86.3\%.

In the paper \cite{masum2022quantum}, the authors investigate the feasibility of QML for detecting Software Supply Chain (SSC) attacks. An SSC attack is characterized by an attacker infiltrating and manipulating a software development or distribution process to inject malware or vulnerabilities into legitimate software. Their comparative study utilized a Quantum Support Vector Machine (QSVM) and  QNN and compared these quantum classifiers with their classical counterparts in terms of speed and accuracy. Due to the access limitations of real quantum computers, they consider open-source quantum simulators, including IBM’s Qiskit and TensorFlow Quantum. The results indicate that both quantum classifiers struggle with accuracy and computation time compared to their classical counterparts. The inferior performance of the quantum classifiers might be attributed to the fact that the simulators do not provide the same capabilities as the original quantum machines. 

In a similar study, the capabilities of QNNs were examined for detecting vulnerabilities in the SSC within source code \citep{Akter10020813}. This comparative study analyzed the performance of QNN and traditional Neural Networks (NN). The experiment was conducted using simulators, with Pennylane employed for the QNN and TensorFlow and Keras utilized for the NN. A software supply chain attack dataset, known as ClaMP, was used to evaluate the performance of both models. Different proportions of the ClaMP dataset were used to identify performance metrics, including F1 score, recall, precision, accuracy, and execution time. The results indicated that both models generally exhibited similar performance, except that the execution time for the QNN was higher than that for the NN. Furthermore, it was observed that the execution time for the QNN slowed down with an increase in the proportion of the dataset, while the execution time for the NN increased with a higher percentage of the dataset.

To enhance the accessibility of QML for individuals with limited expertise,  \cite{amato2023quantumoonlight}  have introduced an online tool called 'QuantumMoonLight'. This software offers a layer of abstraction that masks the inherent intricacies of quantum circuits, granting users a visual interface to engage with quantum machines and implement QML algorithms. The IBM Quantum Computing platform serves as the underlying infrastructure for QML functionalities. Through QuantumMoonLight, users can create, run, and compare QML models. The tool also facilitates various functions, including effortless data loading, data segmentation, preprocessing, feature extraction, feature selection, and QML algorithms. The available algorithms include the QSVM,  Quantum Support Vector Classifier,  QNN Classifier, and the Pegasos QSVC. The efficacy of the tool is assessed through the execution of two software engineering prediction assignments: code smell prediction and flaky test prediction. In both tasks, the QSVM was utilized for model construction. The researchers observed that the QSVM required less training time when compared to traditional machine learning methods in both instances. While the outcomes of the code smell prediction task demonstrated consistent accuracy similar to classical machine learning techniques, the prediction performance of the QSVM was deemed less effective than classical computing techniques in the context of the flaky test prediction task.

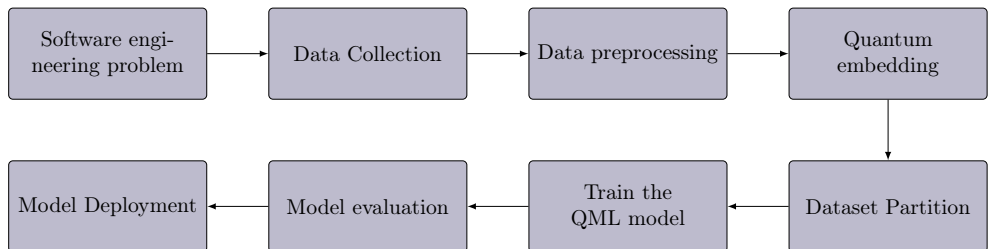
\begin{figure}[!htb]
    \centering
	\tikzset{block/.style={draw, text width=3cm, minimum height=1.5cm, align=center, font=\rmfamily, rectangle,fill=CadetBlue!40,rounded corners=2pt,},
		line/.style={-latex}
	}
	\resizebox{\textwidth}{!}{
	\begin{tikzpicture}
        \node[block] (1) {Software engineering problem};
    \node[block, right=of 1] (2) {Data Collection};
	\node[block,right =of 2] (3)  {Data preprocessing};
	\node[block,right=of 3] (4)  {Quantum embedding };
	\node[block,below=of 4] (5) {Dataset Partition};
	\node[block,left=of 5] (6)  {Train the QML model};
        \node[block,left=of 6] (7)  {Model evaluation};
        \node[block,left=of 7] (8)  {Model Deployment};

	\draw[line] (1) -- (2);
	\draw[line] (2)-- (3);
	\draw[line] (3)--(4);
	\draw[line] (4)-- (5) node[pos=0.3,left]{};
	\draw[line] (5)-- (6) ;
        \draw[line] (6)-- (7) ;
        \draw[line] (7)-- (8) ;
	\end{tikzpicture}
 }
    \caption{The workflow of solving a software engineering problem using QML algorithm.}
    \label{fig:flowchart_qml}
\end{figure}

The discussion above highlights the ongoing exploration of how QML, mainly supervised learning, can enhance software engineering tasks, specifically forecasting. As QML is advancing rapidly, its potentiality can be explored in various software engineering areas, including bug prediction, spam detection, sentiment analysis, code classification, clustering, effort estimation, topic modelling, and software quality prediction. In these areas, classical machine learning has already been utilized. To guide researchers and developers, we have outlined a generalized workflow demonstrating how supervised QML can address basic classical software engineering tasks, as shown in Figure \ref{fig:flowchart_qml}. Initially, a specific problem related to classical software engineering, such as bug prediction or effort estimation, is chosen to be solved using quantum computing. Relevant data for this task is collected and undergoes preprocessing, which includes data cleansing and handling any missing values. This cleaned data is converted into quantum states using amplitude encoding or quantum embedding techniques. The dataset is subsequently divided into subsets for training and validation. Quantum algorithms like QSVM or QNN are used to train the quantum model, allowing it to recognize patterns in the data for prediction. After the training stage, the performance of the quantum model is assessed using the validation data. Ultimately, the trained quantum model is applied to predict or classify new, unseen data, ensuring a consistent process from start to finish.

\subsection{Quantum search and Optimization algorithms for Software Engineering Research}

\subsubsection{Quantum optimization algorithms}

\begin{figure}[!htb]
    \centering 
\tikzset{
    basic/.style  = {draw, text width=3cm, align=center, font=\sffamily, rectangle,fill=CadetBlue!40,rounded corners=2pt,},
    xnode/.style = {basic, thin, rounded corners=2pt, align=center, fill=PineGreen!60,text width=5cm,},
}
\begin{forest} for tree={
    grow=east,
    growth parent anchor=west,
    parent anchor=east,
    child anchor=west,
    edge path={\noexpand\path[\forestoption{edge},->, >={latex}] 
         (!u.parent anchor) -- +(10pt,0pt) |-  (.child anchor) 
         \forestoption{edge label};}
}
[Quantum optimization approaches, basic,  l sep=10mm,
    [Quantum Approximate Optimization Algorithm (QAOA), xnode,  l sep=10mm]
    [Variational Quantum Eigensolver (VQE), xnode,  l sep=10mm]
    [Quantum Gradient Descent, xnode,  l sep=10mm] 
    [Quantum Annealing, xnode,  l sep=10mm]
    ]
\end{forest}
    \caption{Different Quantum optimization approaches}
    \label{fig:quantum_optimum}
\end{figure}
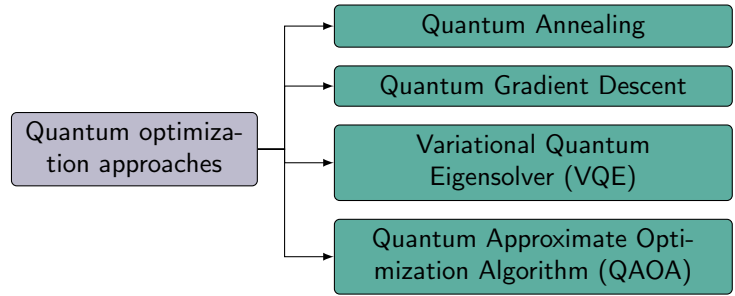

Numerous problems across various domains are computationally challenging to solve on classical machines within a predetermined time. For example, combinatorial optimization problems are computationally expensive because the number of possible combinations grows exponentially with the increase in the number of instances. To manage this computational expense, different strategies such as approximation algorithms, heuristics, meta-heuristic searches, and greedy strategies are often employed to generate quality solutions within a reasonable time. However, these methods usually result in local optima rather than the best solution. To address these types of problems more effectively, quantum optimization algorithms have emerged as an alternative approach. These algorithms leverage principles from quantum mechanics to find the best possible solution and solve optimization problems more efficiently than classical algorithms. Figure \ref{fig:quantum_optimum} lists popular quantum optimization approaches. In addition to Quantum Annealing, three other techniques, the Quantum Approximate Optimization Algorithm (QAOA), Variational Quantum Eigensolver (VQE), and Quantum Gradient Descent, are gate-based optimization methods. We have provided concise summaries of these algorithms in the following.

\begin{itemize}

\item \textit{Quantum annealing}: Quantum annealing is a specialized form of AQC  designed specifically for addressing optimization problems \citep{grant2020adiabatic}. In quantum annealing,   an optimization problem is encoded into an energy landscape represented by a mathematical Hamiltonian. Initially, the quantum system starts with a superposition of potential solutions corresponding to a high-energy state. Through a gradual evolution facilitated by adjustments to the Hamiltonian, the system navigates the energy landscape to reach the lowest energy state. From this energy state, the optimal or near-optimal solution is calculated. Quantum annealing usually employs two mathematical frameworks to reformulate complex optimization problems into a form suitable for quantum processing. These models include the Ising and Quadratic Unconstrained Binary Optimization (QUBO) models. Both models exhibit similarities in their formulation, and their conversion process usually entails a simple linear mapping of variables. The strategy behind quantum annealing is that if an optimization problem can be reformulated into an Ising or QUBO model, a quantum annealer can map this model onto a quantum processor to effectively find the minimum energy solution \citep{heng2022solve}. The quantum computing vendor D-Wave system uses quantum annealing to solve the optimization problem. Here, we have presented briefly how  QUBO mathematical models are formulated for optimization problems. In QUBO, the objective is to find the binary vector $x$ that minimizes the value of the objective function \citep{lewis2017quadratic}. The optimization (minimization) function of QUBO problems is defined as follows:

\begin{equation}
Q(x)=\sum_{i}Q_{i,i}X_i+\sum_{i<j} Q_{i,j} X_iX_j
\end{equation}

where $Q(x)$ is the QUBO objective function, $X=\{X_1,X_2,...,X_n\}$ represents the vector of binary variables and Q is a QUBO matrix, which is an $n \times n$ upper triangular matrix with real weights.  The linear terms are represented by the diagonal of the $Q$ matrix and the quadratic terms are represented by the off-diagonal elements.  The goal is to identify the optimal binary vector configuration by minimizing the QUBO objective function. Figure \ref{fig:flowchart_QA} presents the workflow diagram for quantum annealing.

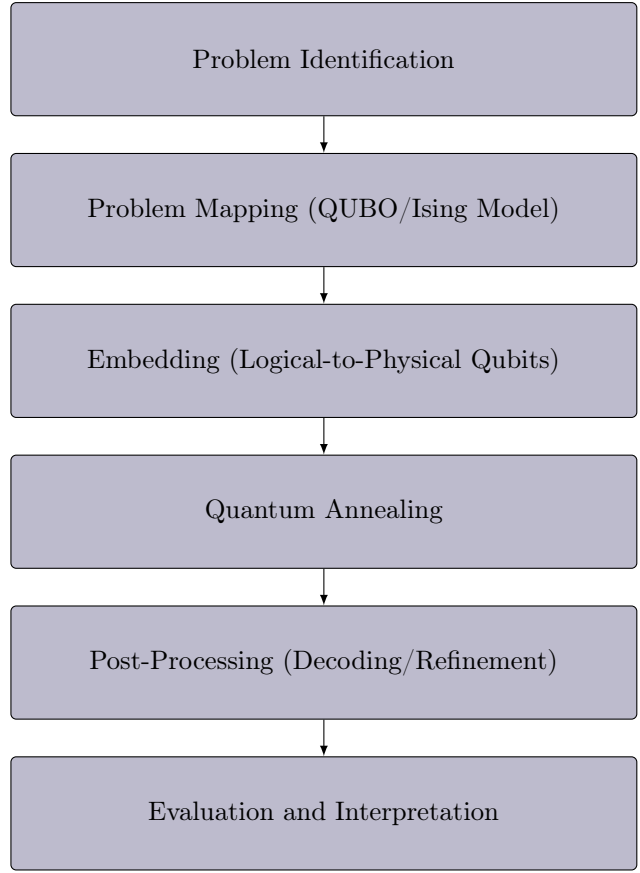
\begin{figure}[!htb]
    \centering
	\tikzset{process/.style={draw, text width=8cm, minimum height=1.5cm, align=center, font=\rmfamily, rectangle,fill=CadetBlue!40,rounded corners=2pt,},
		arrow/.style={-latex}
	}
	\begin{tikzpicture}[node distance=2cm]

       \node (start) [process] {Problem Identification};
        \node (map) [process, below of=start] {Problem Mapping (QUBO/Ising Model)};
        \node (embed) [process, below of=map] {Embedding (Logical-to-Physical Qubits)};
        \node (anneal) [process, below of=embed] {Quantum Annealing};
        \node (post) [process, below of=anneal] {Post-Processing (Decoding/Refinement)};
        \node (evaluate) [process, below of=post] {Evaluation and Interpretation};
        
        \draw [arrow] (start) -- (map);
        \draw [arrow] (map) -- (embed);
        \draw [arrow] (embed) -- (anneal);
        \draw [arrow] (anneal) -- (post);
        \draw [arrow] (post) -- (evaluate);
        
        \end{tikzpicture}
            \caption{ The workflow diagram for a quantum annealing process}
            \label{fig:flowchart_QA}
\end{figure}

\item \textit{Variational Quantum Eigensolver (VQE)}: VQE \citep{peruzzo2014variational} is a hybrid algorithm that combines quantum and classical computing principles to address complex optimization problems. Initially designed for quantum chemistry simulations, VQE can also be adapted for optimization tasks. Its primary goal is to find the smallest eigenvalue of a Hermitian matrix, such as determining the ground state energy. VQE uses a Parameterized Quantum Circuit (PQC) to create a quantum state, which is then used to calculate the expectation value of a given Hamiltonian \citep{sim2019expressibility}. The algorithm operates iteratively, adjusting the parameters to minimize the expected energy. In each iteration, the quantum circuit prepares an ansatz, or trial state, and evaluates its energy. A classical optimizer then updates the parameters of the ansatz to minimize the energy value further. It is well-suited for current NISQ devices. Figure \ref{fig:vqe} illustrates the VQE algorithm.

Let the problem is formulated as a Hamiltonian \( H \), and a trial wavefunction \( |\psi(\theta)\rangle \) is prepared on a quantum computer using a parameterized quantum circuit \( \hat{U}(\theta) \). The circuit parameters \( \theta \) are iteratively adjusted to approximate the ground state of the Hamiltonian. The expectation value of the Hamiltonian is calculated for the current parameter values as:

\[
E(\theta) = \langle \psi(\theta) | H | \psi(\theta) \rangle.
\]

Due to the probabilistic nature of quantum mechanics, repeated measurements are performed to estimate \( E(\theta) \). A classical optimizer then minimizes this expectation value by updating the parameters \( \theta \) until convergence.

\begin{figure}[!htb]
    \centering
    \includegraphics[width=0.5 \linewidth]{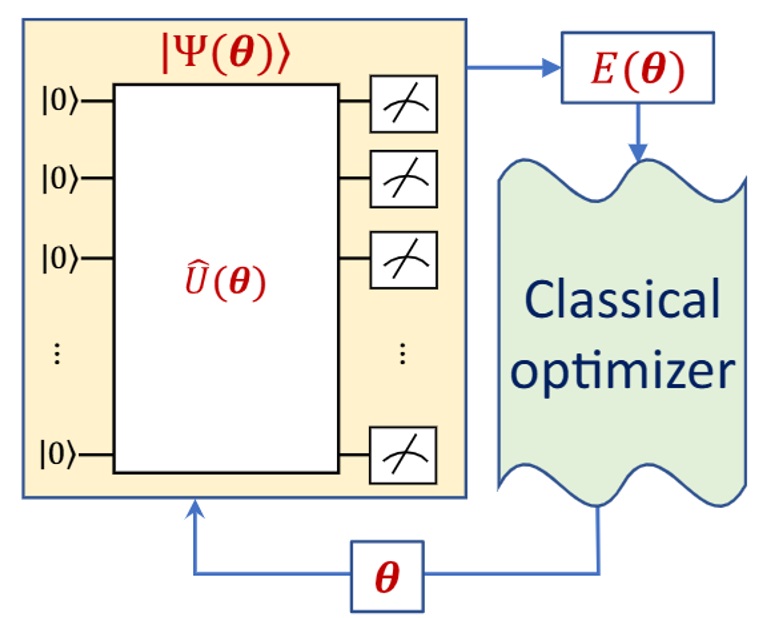}
    \caption{A schematic representation of the VQE algorithm \citep{Mukherjee2023}}
    \label{fig:vqe}
\end{figure}

\item \textit{Quantum Approximate Optimization Algorithm (QAOA)}: The QAOA \citep{Choi8939749} is another popular hybrid classical-quantum algorithm specifically designed to address optimization problems. It can be viewed as a specialized variant of the VQE algorithm, with a focus on combinatorial optimization problems. The goal of QAOA is to identify parameters that optimize the expectation value of a cost function associated with the problem at hand. Like VQE, QAOA demonstrates a level of resilience to noise, making it suitable for use on NISQ devices \citep{Lavrijsen9259985}. While QAOA and VQE share some similarities, QAOA sets itself apart by employing a sequence of two-parameter unitary operations to construct a more complex ansatz (trial quantum state). QAOA seeks to find approximate solutions to intricate optimization problems by initially mapping them onto a Hamiltonian. It then builds a PQC that begins with qubits in a superposition state. This circuit consists of alternating layers representing the problem and a simpler mixer operation, both of which have tunable parameters. The results are then measured, and a classical optimizer adjusts the parameters based on these measurements. This iterative process of circuit construction, measurement, and classical optimization guides the quantum state toward a configuration that represents an optimal solution to the original problem. 

Figure \ref{fig:QAOA}  illustrates the basic workflow of the Quantum Approximate Optimization Algorithm (QAOA). The optimization problem is encoded into a cost Hamiltonian \( C \). The parameterized quantum circuit alternates between applying the cost Hamiltonian \( C \) and a mixing Hamiltonian \( B \) for \( p \) layers. The quantum state, controlled by parameters \( \vec{\beta} \) and \( \vec{\gamma} \), evolves to minimize the expectation value:
\[
C_p = \bra{\psi(\vec{\beta}, \vec{\gamma})} \mathcal C \ket{\psi(\vec{\beta}, \vec{\gamma})}.
\]
A classical optimizer iteratively updates \( \vec{\beta} \) and \( \vec{\gamma} \) to converge towards the optimal solution.

\begin{figure}[!htb]
    \centering
    \includegraphics[width=0.9\linewidth]{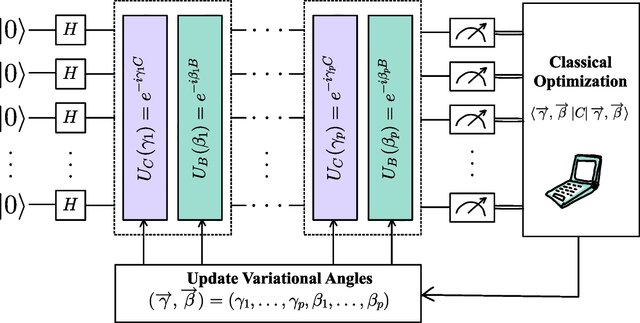}
    \caption{Schematic Representation of the QAOA \citep{Falla2024} }
    \label{fig:QAOA}
\end{figure}

\item \textit{Quantum Gradient Descent (QGD)}:  QGD \citep{PhysRevA.101.022316} is  quantum adaptation of the traditional gradient descent algorithm. It leverages quantum operations to compute gradients and update parameters, potentially offering a speedup for certain optimization tasks. QGD is specifically designed to work with PQCs. The objective is to fine-tune the parameters of these circuits to achieve the optimal result. The primary advantage of the algorithm is its ability to harness quantum parallelism, which allows for the simultaneous evaluation of multiple gradient components. This parallelism can lead to faster convergence compared to classical gradient-based methods.
\end{itemize}

\subsubsection{Applications of Quantum Optimization in Software Engineering Research}

In this subsection, we have discussed the application of quantum optimization and quantum search algorithms in software engineering, highlighting their potential to expedite various tasks within the field.

\cite{Miranskyy2022} introduced a quantum search strategy to speed up exhaustive and non-exhaustive dynamic testing methods designed for classical computers. They combined Grover’s search algorithm with a quantum counting algorithm to expedite the acquisition of testing results. They demonstrated that the application of a quantum algorithm allows for a decrease in computational complexity from $O(N)$ to $O(\epsilon^{_1}\sqrt{N/K})$, where  $K$ is number of inputs causing errors, with $\epsilon$ is a relative error of measuring $k$,  $N$ represents the count of combinations of input parameter values passed to the software under test. Their experimental study utilized a simple example that was run on a simulator and an actual quantum computer. The main hurdle to improving the practicality of this method is the refinement of the conversion process from classical to quantum circuits and the enhancement of the speed, size, and resilience of quantum computers.

Identifying the minor test suite that encompasses all requirements poses a significant challenge for traditional computing, mainly when the number of test cases is considerably large. \cite{hussein2021quantum} have proposed a quantum search algorithm to accelerate the search process and identify the optimal test suite in software engineering. This algorithm can find the optimal test suite with a high probability in $O(\sqrt{2^n})$, where $n$ is the number of test cases. The algorithm employs amplitude amplification techniques to search for the minimum test cases to cover all requirements. The proposed algorithm is composed of two stages. Initially, in the preparation phase, it implements the quantum searching methodology proposed by  \cite{YOUNES20081074}, which harnesses principles of entanglement to establish a prepared superposition. Subsequently, in the search phase, the algorithm integrates Arima's algorithm \citep{5369930} to explore potential solutions within the established superposition. Although a comprehensive algorithm description has been provided, no experimental analysis using quantum computing or simulation has yet been conducted.


In software engineering research, identifying code clones is a crucial task as it assists in enhancing the quality of code, minimizing duplication, enhancing maintainability, and preventing potential bugs in software projects. Code clones are the sections of code within a software project that are identical or highly similar. The taxonomy of code clones encompasses four main types: Exact Clones (Type 1), Renamed Clones (Type 2), Near Miss Clones (Type 3), and Semantic Clones (Type 4).  \cite{Jhaveri2023} presented how a D-Wave quantum computer can be utilized for code clone detection. In this study, particular emphasis was placed on detecting Type 3 clones due to their inherent complexity and computational demands. To address the problem, they formulate code clone detection as a subgraph isomorphism problem, which is then transformed into a quadratic optimization problem. This problem is subsequently solved using a D-Wave quantum annealing computer. The authors first represent programs as graphs using Abstract Syntax Trees (ASTs) in their methodology. To facilitate code clone detection between a target graph (G1) and a query graph (G2), subgraph isomorphism algorithms, inspired by  \cite{zick2015experimental}, were employed. These problems were then transformed into Quadratic Unconstrained Discrete Optimization models (QUDO), which were automatically converted into QUBO formulations. To precisely detect code clones, they consider node types penalty, node-mapping penalty, and edge-mapping penalty. The QUBO problems are solved using quantum annealing, and decisions about the code clone percentage are made based on the global minimum energy state. Their exploratory study suggests that quantum annealing holds promise for detecting Type 2 and Type 3 code clones. However, their study is based on a small case, and a comparative performance analysis with other studies is required for a complete evaluation.

Various software engineering applications, such as formal software verification \citep{ivanvcic2008efficient}, static analysis, software model verification, false path pruning \citep{yang2009dynamic}, and test suite reduction, have extensively utilized the Boolean Satisfiability (SAT) problem. The challenge of SAT is to find if there is a set of truth values (either true or false) for the variables that would make a given Boolean formula accurate. Given the computational difficulty of SAT, quantum optimization algorithms have shown promise in providing efficient solutions. For example, D-Wave quantum annealing has been effectively used to tackle such problems \citep{kruger2020quantum, bian2020solving}. In addition, IBM QAOA has proven to be effective in solving 3-SAT and Max-3-SAT problems \citep{yu2023solution}.

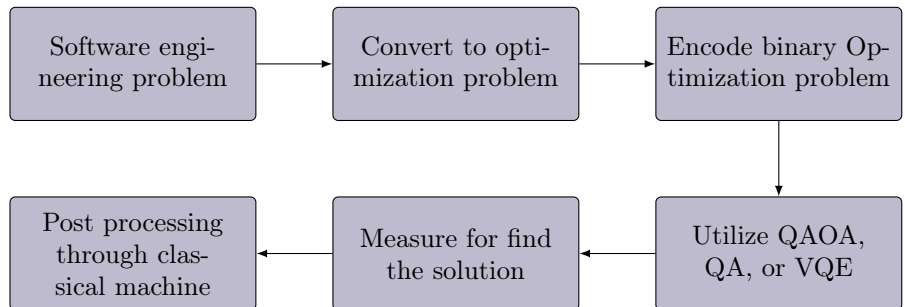
\begin{figure}[!htb]
    \centering
	\tikzset{block/.style={draw, text width=3cm, minimum height=1.5cm, align=center, font=\rmfamily, rectangle,fill=CadetBlue!40,rounded corners=2pt,},
		line/.style={-latex}
	}
	
	\begin{tikzpicture}
        \node[block] (1) {Software engineering problem};
    \node[block, right=of 1] (2) {Convert to optimization problem};
	\node[block,right =of 2] (3)  {Encode binary Optimization problem };
	\node[block,below=of 3] (4)  {Utilize QAOA, QA, or VQE  };
	\node[block,left=of 4] (5) {Measure for find the solution};
	\node[block,left=of 5] (6)  {Post processing through classical machine};
	
	\draw[line] (1) -- (2);
	\draw[line] (2)-- (3);
	\draw[line] (3)--(4) node[pos=0.3,left]{};
	\draw[line] (4)-- (5);
	\draw[line] (5)-- (6) ;
	\end{tikzpicture}
    \caption{The workflow of solving a software engineering problem using quantum optimization algorithm.}
    \label{fig:flowchart_qo}
\end{figure}

Due to its potential to tackle complex computational problems faster than classical computing, the quantum optimization Algorithm has been gaining popularity in many application areas. Although Quantum Optimization and search algorithms have been applied in only a few software engineering research areas, they have great potential to tackle various other software engineering challenges involving optimization problems effectively. These include software project management, code optimization, feature selection, resource allocation, and scheduling. To assist eager researchers in this field, we have illustrated the workflow that shows how quantum optimization algorithms can be employed to solve important software engineering problems, as depicted in Figure \ref{fig:flowchart_qo}. First, the problem is assessed to determine if it can be transformed into a combinatorial optimization problem. If suitable, the objective function is encoded in a binary optimization form compatible with quantum computation. Any Quantum Optimization Algorithms, such as QAOA, quantum annealing, or VQE, are then employed to solve the problem. These algorithms utilize quantum circuits or annealing hardware to explore the solution space, evolving quantum states that minimize the objective function. The obtained quantum states are measured or sampled to extract approximate solutions, which are further refined or validated through classical post-processing techniques.

\subsection{Case Study: Software Bug Prediction}
We conducted a case study to evaluate the performance of classical Random Forest (RF) \citep{Scikit:learnML} and Quantum Random Forest (QRF) \citep{Srikumar2024} for software defect prediction (SDP) using 10 datasets from the PROMISE repository \citep{Sayyad-Shirabad+Menzies:2005}. This repository is widely recognized in SDP research, and we selected 10 software projects from the AEEM and JIRA categories, as detailed in Table \ref{tab:datasets-promise}. The datasets contain an imbalanced number of data instances in buggy and non-buggy software instances. To address the data imbalance of these software projects, we applied the Synthetic Minority Oversampling Technique (SMOTE) \citep{SMOTECite} before implementing the RF and QRF algorithms to predict buggy software instances. The experiment was conducted iteratively, starting with a maximum of 100 data instances per software project and incrementing the total number of instances by 100 in each iteration, up to a maximum of 500 data instances from each software project. We limited the maximum data instance from each software project to 500 because of the high runtime demand of the QRF algorithm. QRF transforms the classical feature value to its quantum feature map, which takes a considerable amount of run time (ranging from 500 to 2000 seconds). To avoid slower processing of the experiments of this case study, we applied the limit of the maximum number of data instances from each software project. As we incremented the data instances by 100 in each iteration, starting from the initial 100, it resulted in five iterations for each dataset. Such an iterative process enables us to perform a comparative evaluation of RF and QRF across varying dataset sizes. This iterative approach demonstrated the practical applicability of the QRF algorithm in real-world scenarios, highlighting its effectiveness compared to RF under different data conditions (software project and dataset size).

\begin{table}[htb!]
	\centering
	\caption{Software Defect Prediction Datasets}
	\label{tab:datasets-promise}
	\begin{tabular}{lcccc}
		\toprule
		\textbf{Projects} & \textbf{Dataset} & \textbf{Features} & \textbf{Samples} & \textbf{Class Distribution} \\ 
		\midrule
		\multirow{5}{*}{AEEEM} 
		& EQ       & 39 & 324  & \{0: 195, 1: 129\}  \\ 
		& JDT      & 43 & 997  & \{0: 791, 1: 206\}  \\ 
		& Lucene   & 40 & 691  & \{0: 627, 1: 64\}   \\ 
		& Mylyn    & 38 & 1862 & \{0: 1617, 1: 245\} \\ 
		& PDE      & 39 & 1497 & \{0: 1288, 1: 209\} \\ 
		\midrule
		\multirow{5}{*}{JIRA} 
		& Activemq & 44 & 1884 & \{0: 1591, 1: 293\} \\ 
		& Groovy   & 44 & 821  & \{0: 751, 1: 70\}   \\ 
		& HBase    & 43 & 1059 & \{0: 841, 1: 218\}  \\ 
		& Hive     & 42 & 1416 & \{0: 1133, 1: 283\} \\ 
		& Wicket   & 45 & 1763 & \{0: 1633, 1: 130\} \\ 
		\bottomrule
	\end{tabular}
\end{table}

\subsubsection{Dataset Size: 100}
\begin{figure}
	\centering
	\includegraphics[width=\textwidth] {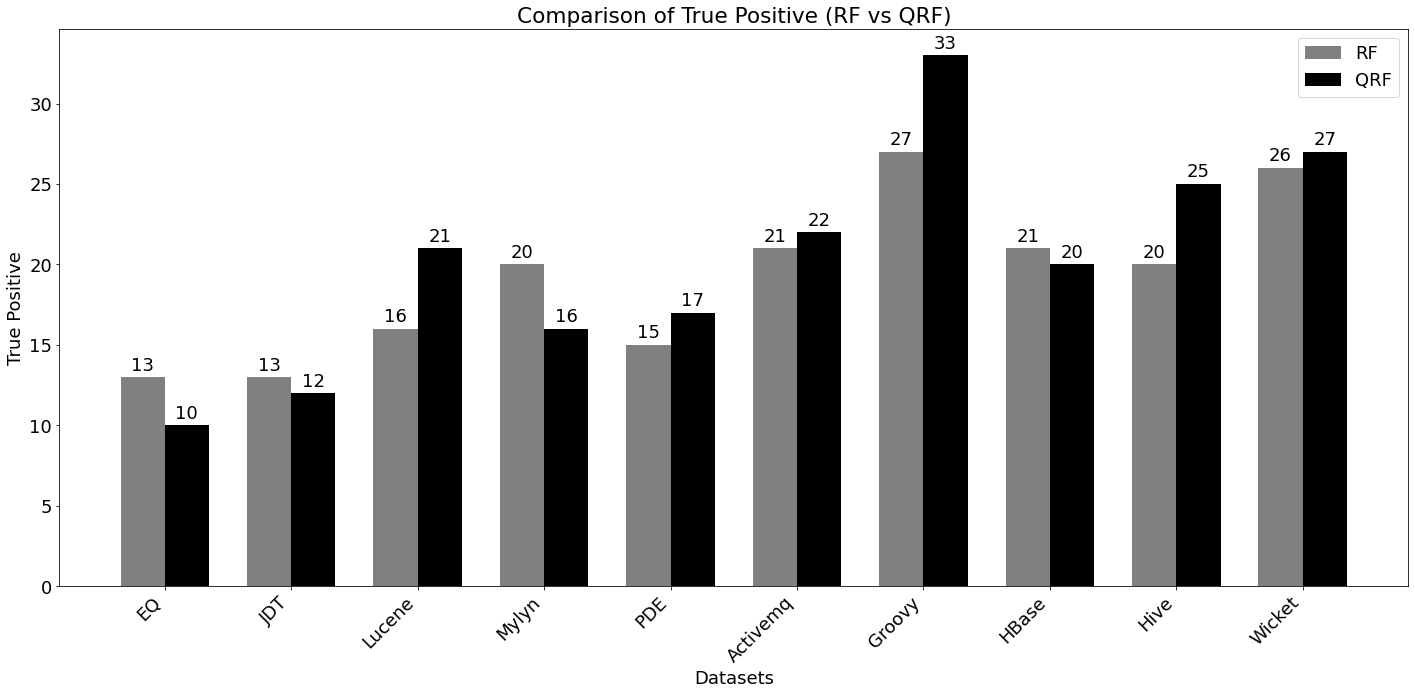}
	\caption{Comparing True Positives (TP) in Bug Prediction}
	\label{fig:compTP}
\end{figure}

When selecting 100 data instances from each subject system, we observed that the classical Random Forest (RF) struggled to outperform the Quantum Random Forest (QRF) in terms of true positives (TP). True positives refer to the number of correctly predicted buggy software instances, highlighting the effectiveness of a predictive model. Figure \ref{fig:compTP} compares the TP predictions of RF and QRF, showing that QRF outperformed RF in six out of ten software projects. In the remaining four projects, QRF achieved TP values comparable to RF. These findings underscore the advantages of using QRF over RF in scenarios with limited data, where classical algorithms may lack sufficient training to perform effective prediction of buggy data instances. Our case study results suggest that in data-scarce environments, quantum algorithms like QRF offer a significant performance advantage over their classical counterparts in such a difficult prediction scenario.

\begin{figure}
	\centering
	\includegraphics[width=\textwidth] {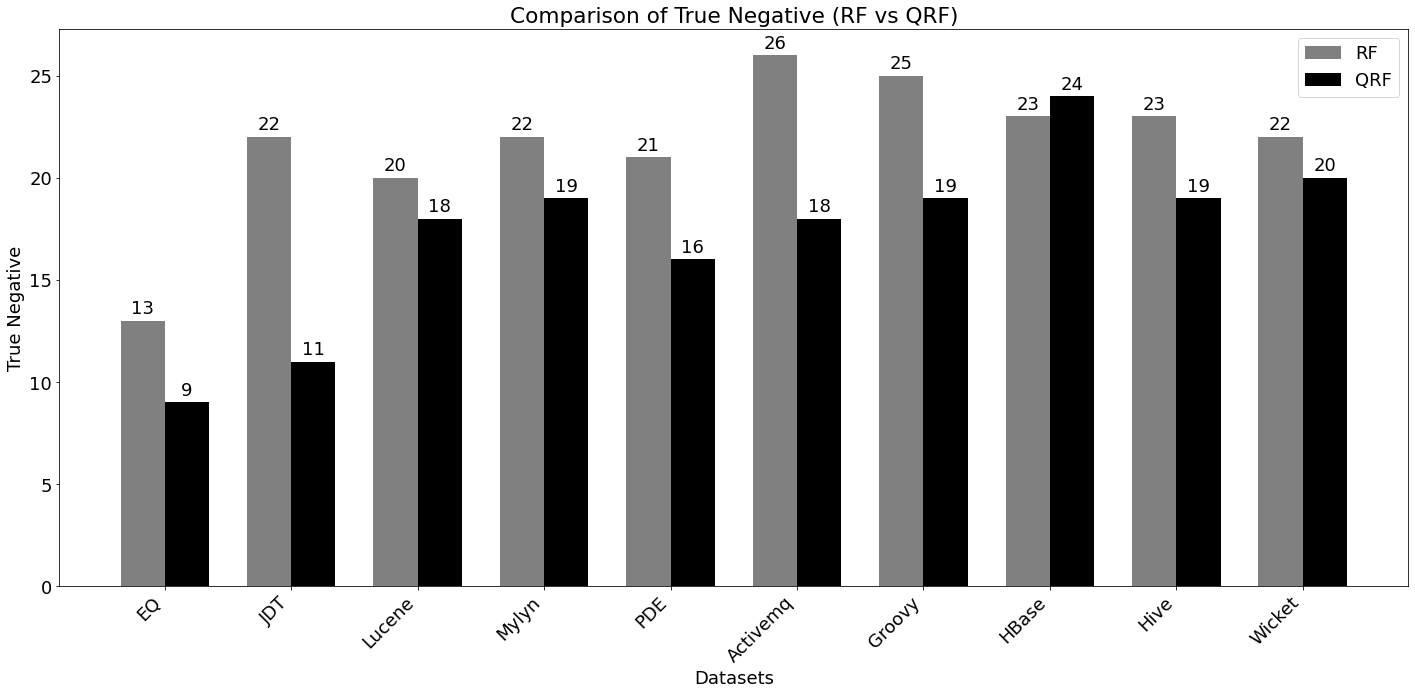}
	\caption{Comparing True Negative (TN) in Bug Prediction}
	\label{fig:compTN}
\end{figure}

\begin{figure}
	\centering
	\includegraphics[width=\textwidth] {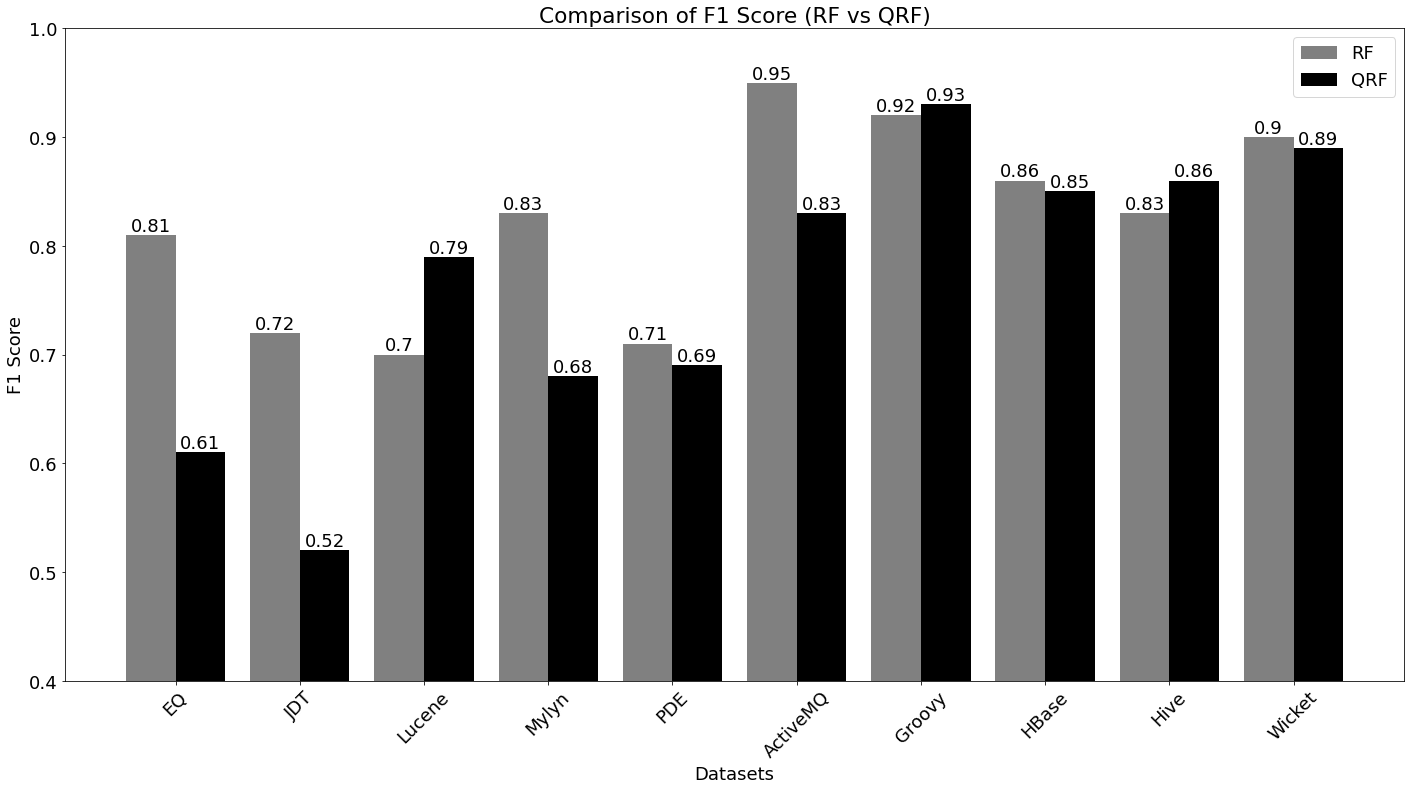}
	\caption{Comparing F1 Score in Bug Prediction}
	\label{fig:compF1}
\end{figure}

We compare the F1-scores for predicting both buggy and non-buggy software instances in Figure \ref{fig:compF1}, illustrating how the performance metrics differ in detecting both buggy and non-buggy data instances. F1~Score is the harmonic mean of precision and recall values calculated using the True Positive (TP), True Negative (TN), False Positive (FP), and False Negative (FN) values, as the following formula. 

\[ Precision = \frac{TP}{TP + FP}\]
\[ Recall = \frac{TP}{TP + FN}\]
\[ F1~Score = \frac{2 \times Precision \times Recall}{Precision + Recall}\]

Although the TP values in Figure \ref{fig:compTP} suggest that QRF generally outperforms RF across most software projects, QRF demonstrates limited effectiveness in detecting True Negatives (TN), as illustrated in Figure \ref{fig:compTN}. This limitation is also evident in the F1-score differences shown in Figure \ref{fig:compF1}. Notably, the F1-score variations between RF and QRF across different software projects in Figure \ref{fig:compF1} are relatively minor. This observation suggests that although QRF outperforms at identifying buggy instances correctly, it struggles to effectively capture non-buggy instances, which is the reason for degrading the F1~Scores. However, detecting buggy instances is more critical than non-buggy ones due to the significant impact that undetected bugs can have on software projects. Consequently, the choice of using QRF in real-world software development depends on the specific prediction objective. If prioritizing the accurate detection of buggy instances is critical, the Quantum algorithm is preferable, particularly when the dataset is too small to train a classical ML algorithm effectively.

\subsubsection{Iterating the Dataset Size: 100 to 500}
\begin{figure}
	\centering
	\includegraphics[width=\textwidth] {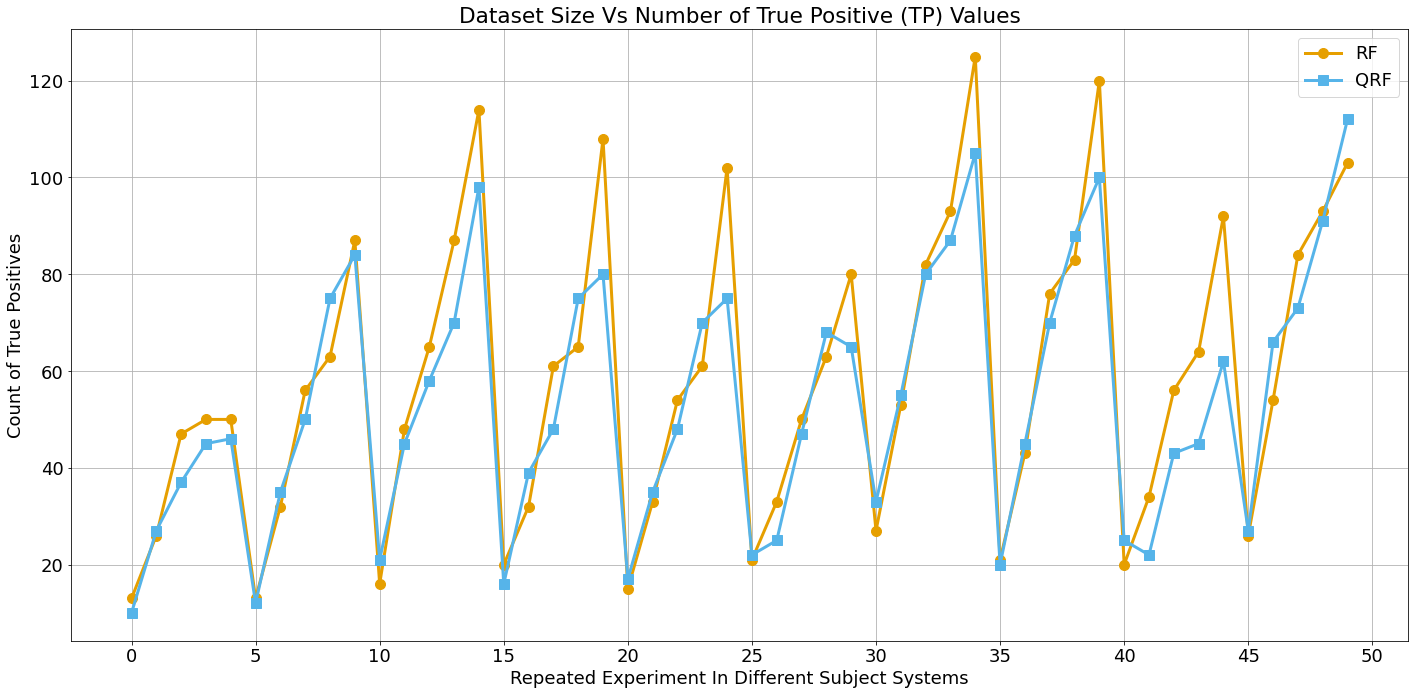}
	\caption{Comparing TP across Different Dataset Sizes: Each 5-Unit Interval (e.g., 0-5, 5-10, etc.) Represents a Distinct Subject System.}
	\label{fig:TPChange}
\end{figure}

In this step, we conducted the experiment five times, beginning with a dataset size of 100 and increasing the size by increments of 100 in each subsequent iteration. The performance of Random Forest (RF) and Quantum Random Forest (QRF) was evaluated across varying dataset sizes derived from different software projects. The comparative results are presented in Figure \ref{fig:TPChange}, which illustrates how the number of True Positives (TP) varies with dataset size across multiple software projects. 

The results for the different datasets in Figure \ref{fig:TPChange} are presented in the same order as in Table \ref{tab:datasets-promise}. For instance, the first group of results, marked as 0 to 5 towards X-axis, corresponds to the software project EQ, and the results of its repeated experiments taking 100 to 500 data instances. The subsequent groups of results are from JDT (5 to 10), Lucene (10 to 15), Mylyn (15 to 20), PDE (20 to 25), and others. A consistent trend is observed in the comparison scenarios depicted in Figure \ref{fig:TPChange}. Specifically, when the dataset size for a software project is smaller (e.g., 100 or 200 instances), QRF outperforms RF in predicting buggy software instances. However, as the dataset size increases (e.g., 400 or 500 instances), RF demonstrates better predictive performance compared to the QRF.

The performance comparison of QRF and RF with the varied size of datasets shows the trade-off of data availability with the feasibility of choosing between these two algorithms. When the dataset size is smaller, it is wise to choose QRF as it utilizes its quantum computing principles in representing classical data to its quantum feature space and finds a better classifying path (decision trees). On the other hand, dataset scarcity makes the classical RF poorly trained and it fails to perform as QRF. When there is enough dataset to train the classical RF algorithm properly, using QRF in such a case is not feasible. We assume that, as the dataset size increases, QRF might become prone to overfitting making the model complexity non-favorable to classify a test dataset effectively. We plan to explore this direction of the study in the future with some different types of datasets and Quantum Machine Learning (QML) algorithms utilized in similar research \citep{NadimQSE2025, AshisQAI2024, QMLQiskitPY, QMLInro}. Another research direction from this can also emerge to investigate the optimal resource requirements for running QML algorithms. QML algorithms might face challenges with scalability and may require more computational resources to maintain their performance as the dataset grows, which could lead to less efficient performance compared to classical models like RF in larger datasets. We also plan to extend our research direction in this direction in the future.

\section{QSE and its progress}

With its immense computational power, quantum computing has the potential to revolutionize various areas. However, harnessing this power necessitates the development of specialized quantum software. QSE provides an array of principles, ideas, and guidelines essential to developing, maintaining, and growing quantum software solutions. The practice of quantum software development is a critical factor in unlocking the full potential of quantum computing. Although it is a relatively new field, QSE has experienced rapid advancements. The first International Workshop on QSE and Programming (QANSWER) has been organized to stimulate further growth and expansion in this area. This workshop has presented a comprehensive set of principles and commitments that aim to guide the future of quantum software development \citep{piattini2020talavera}. In addition, it has given the roles of various stakeholders to provide a roadmap for the successful evolution of this field. 
 \cite{Miranskyy2019} emphasize that as quantum computing becomes more prevalent, the software engineering community needs to begin integrating software engineering practices into the quantum computing domain. The paper outlines the challenges associated with testing, verification, and validation in quantum computing. It highlights that not all existing software engineering (SE) practices are readily transferable to the quantum computing (QC) domain, necessitating the development of new practices where adaptation is complex. For instance, while code reviews are easily transferable to the QC domain, interactive debugging poses significant challenges in terms of transferability. Additionally, the complexity-dependent placement of validation programs requires introducing novel approaches.  \cite{moguel2020roadmap} discusses the current state of quantum programming tools, stating that they are at a primitive level. This presents challenges in developing quantum software systems, similar to the difficulties experienced in the 1960s for classical computing. The paper predicts future improvements in quantum computing and emphasizes the necessity of QSE. Research in this field should focus on creating user-friendly tools for designing and building quantum programs and draw upon classical software engineering practices.

\begin{figure}[!ht]
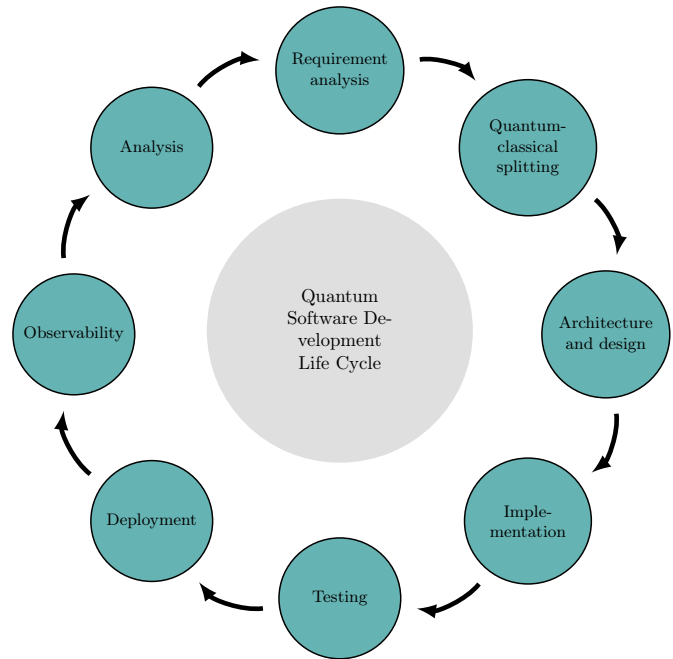

    \centering
    \scalebox{0.7}{ 
    \tikzset{
        every shadow/.style={
        fill=none,
        shadow xshift=0pt,
        shadow yshift=0pt}
    }
    \tikzset{module/.append style={top color=teal!60,bottom color=teal!60}}
    \smartdiagramset{custom/.style={
    arrow tip=latex,
    arrow color=black,
    arrow line width=2.5pt,
    module shape=circle,
    uniform color list=green!50 for 8 items, 
    uniform arrow color=true, 
    text width=2cm,
    circular distance=5cm,
    border color=black,
    additions={
        additional item font=\normalsize,
        additional item fill color=lightgray!50,
        additional item offset=1.20cm,
        additional item text width=2cm,
        additional item width=5cm,
        }
    }
}

\smartdiagramset{custom}
\smartdiagramadd[circular diagram:clockwise]
    { Analysis, Requirement analysis, 
    Quantum-classical splitting, Architecture and design, Implementation, Testing, Deployment, Observability}
    {below of module2/Quantum Software Development Life Cycle }
}
    \caption{A  Quantum Software Development Lifecycle proposed in \citep{Weder2022} }
    \label{fig:sdlc}
\end{figure}

Like classical software development life cycles, quantum software development follows a structured process to design, implement, and deploy quantum algorithms to solve complex computational problems. However, due to the unique characteristics of quantum computing, such as superposition, entanglement, and quantum noise, the development life cycle of quantum software introduces specific considerations and challenges.  
\cite{Weder2020} highlight the importance of improving quantum algorithms and the software tools needed to execute quantum machines effectively. They suggested a quantum software lifecycle with ten phases that gate-based quantum applications should follow.   \cite{Scheerer2023} suggest a phased approach for developing quantum algorithms. The steps of this process encompass defining the problem, choosing the quantum algorithm, implementing the algorithm, fine-tuning, and backpropagation. This model, designed from their hands-on experience, could be a practical initial guide for those developing quantum algorithms. Besides,  \cite{Weder2022} also provides eight phases for the quantum software development lifecycle: Requirement Analysis, Quantum-Classical Splitting, Architecture and Design, Implementation, Testing, Deployment, Observability, and Analysis. They underscored that since most quantum applications are hybrid, the lifecycle for these applications necessitates the development of both types of programs. They proposed an eight-phase framework for the quantum software development process, including requirement analysis, quantum-classical splitting, architecture and design, implementation, testing, deployment, observability, and analysis, as shown in Figure \ref{fig:sdlc}.

Several research papers have been published in recent years that have contributed to the development of QSE. Table \ref{tab:se_practice} summarises key methodologies and tools in QSE to support the development of quantum applications. We have collected the related literature and grouped them into eight different major QSE areas, including Quantum Testing and Debugging, Quantum Formal Methods, Quantum Empirical Software Engineering, Quantum Software Modeling, Benchmarking, Quantum  Requirement Engineering, Quantum  Refactoring, and Quantum  Reverse Engineering.

\begin{longtable} {|p{2cm}|p{7.5cm}|p{2cm}|}
\caption{Practices of QSE in Literature}
\label{tab:se_practice} \\

\hline
\textbf{Area} & \textbf{Description} & \textbf{References} \\
\hline
\endfirsthead

\multicolumn{3}{c}{{\bfseries \tablename\ \thetable{} -- continued from previous page}} \\
\hline
\textbf{Area} & \textbf{Description} & \textbf{References} \\
\hline
\endhead

\hline \multicolumn{3}{|r|}{{Continued on next page}} \\ 
\hline
\endfoot

\hline 
\endlastfoot

Quantum Testing and Debugging & A benchmark suite named Bugs4Q featuring thirty-six manually validated bugs found in popular Qiskit programs. It also provides a method to access and execute test cases on the error-prone and corrected versions of Qiskit programs. This feature assists in conducting reproducible research studies and comparing tools utilized for debugging and testing the Qiskit program. & \citep{Zhao9678908} \\ 
\hline

Quantum Testing and Debugging & Automated techniques and tools to recognize, examine, and detect bug-fix patterns for quantum software. The paper introduces Q-Diff, a framework based on Abstract Syntax Trees (AST), designed to automatically identify bug-fix patterns in IBM Qiskit code. This approach is validated by testing Q-Diff with a diverse set of quantum bug-fix patterns. & \citep{kher2023automatic} \\ 
\hline

Quantum Testing and Debugging & Quito (Quantum Input Output coverage), a paradigm developed to automate and systematically test quantum programs. It includes three coverage criteria explicitly designed for the inputs and outputs of a quantum program and related strategies for test generation. Effectiveness is assessed through mutation analysis and experimental trials on five different quantum programs. & \citep{ali2021assessing} \\
\hline

Quantum Testing and Debugging & Quito, a quantum software testing tool, is developed to create test suites that meet input coverage, output coverage, and input-output coverage criteria. These criteria are defined based on the inputs and outputs of a quantum program written in Qiskit. Quito can automatically verify the accuracy of the quantum program using generated tests and employing two kinds of test oracles. & \citep{Wang2021Quito} \\
\hline

Quantum Testing and Debugging & QuSBT, a tool for testing quantum programs, utilizes a genetic algorithm to automate the generation of test cases. Its primary goal is constructing an exhaustive test suite by maximizing the identification of failed test cases, using IBM's Qiskit as a simulation framework for quantum programs. & \citep{Wang2022} \\
\hline

Quantum Testing and Debugging & QSharpCheck, a property-based testing framework for Q\# quantum programs. It specifies property specification, test case generation, and test result analysis within the framework. & \citep{Honarvar2020} \\
\hline

Quantum Testing and Debugging & QMutPy, a tool designed for automated mutation testing of Qiskit programs. Muskit is another similar tool for analyzing Qiskit quantum program mutations. Both tools assess the quality and reliability of test suites for quantum programs, aiding developers in identifying potential weaknesses and enhancing the overall quality of their quantum code. & \citep{Fortunato2022} \citep{Mendiluze2021} \\
\hline

Quantum Testing and Debugging & A survey highlighting various quantum software testing approaches available in the literature: coverage criteria, differential testing, fuzz testing, mutation testing, and metamorphic testing. For instance, Qdiff focuses on differential testing of quantum software stacks, while metamorphic testing addresses the well-known quantum measurement problem. & \citep{Ali2023} \\
\hline

Quantum Software Modelling & A conceptual model proposed for quantum programs written in Microsoft Q\# and IBM Qiskit. An example of modeling a quantum entanglement program is also presented. These models support the generation of quantum codes, as well as quantum verification and validation. & \citep{ali2020modeling} \\
\hline

Quantum Software Modelling & As classical UML is insufficient for quantum software designs, this paper suggests using existing UML extensions for quantum software design instead of starting from scratch. Five guiding principles behind any Q-UML (Quantum Unified Modeling Language) are practiced. An example is a Q-UML class diagram and sequence diagram for Shor’s Algorithm. & \citep{perez2020towards} \\
\hline

Quantum Software Modelling & A UML extension proposed to model quantum circuits as activity diagrams. Creating quantum UML profiling allows for the integrated modeling of both classical and quantum aspects in design. & \citep{Luis9474570} \\
\hline    

Empirical Software Engineering & An empirical study employing Interpretive Structure Modeling (ISM) and Fuzzy Technique for Order of Preference by Similarity to Ideal Solution (Fuzzy TOPSIS) methodologies to systematically identify challenges in the QSE process. Twenty-two obstacles were discovered and grouped into seven main categories, with "complex programming," "limited software libraries," "maintenance complexity," "lack of training and workshops," and "data encoding issues" among the primary barriers. & \citep{Akbar2023} \\
\hline              

Empirical Software Engineering & An empirical study investigating difficulties faced by developers in the quantum software field. The study involved analyzing discussions on Stack Exchange forums related to QSE and examining issue reports from quantum computing projects on GitHub. Developers in QSE face conventional software engineering challenges and quantum-specific challenges, such as deciphering quantum program execution results. & \citep{aoun9609196} \\
\hline

Empirical Software Engineering & A comprehensive empirical analysis of bugs found in quantum computing platforms. Examining 223 real-world bugs from 18 open-source quantum computing platforms, the research focuses on quantum-specific bugs' quantity, locations, manifestations, recurring patterns, and complexity. It introduces new patterns specific to quantum bugs. & \citep{Paltenghi2022} \\
\hline

Empirical Software Engineering & An empirical study examining 391 real bugs from 22 open-source databases spanning nine well-known QML platforms to identify errors in QML systems. One-quarter of these bugs are specific to quantum computing. The most common symptom is a system crash, followed by functional errors. Algorithm and logic errors account for most bugs, with inconsistent behavior being the second most common cause. & \citep{Zhao2023} \\
\hline

Empirical Software Engineering & Semi-structured interviews with 26 professionals from 10 countries to determine the suitability of agile methodologies for quantum software development. The Grounded Theory was employed to extract knowledge from the interviews. Results indicate agile methodologies are well-suited for quantum software development. & \citep{khan10234254} \\
\hline
  
Empirical Software Engineering & A study focused on detecting bugs in quantum circuits introduced by developers, compiled from sources such as StackExchange, StackOverflow, and pertinent research papers. Bugs are classified into categories like initialization errors, gate order issues, and quantum logic problems. & \citep{10313647Metwalli} \\
\hline

Quantum Formal Method & Utilizing the Z formal specification language from classical software engineering, a formal quantum software engineering (F-QSE) method was proposed. F-QSE can rigorously analyze and verify the correctness of quantum algorithms, with a verification example of the Deutsch Algorithm. & \citep{Cartiere2022} \\
\hline

Quantum Formal Method & A survey introducing several techniques and tools for the formal verification of quantum programs to mitigate errors and bugs. Methods include quantum weakest preconditions, quantum Hoare logic, quantum computation tree logic, and ZX-Calculus Path. Notable tools are QWire, SQIR (Small Quantum Intermediate Representation), QHLProver, CoqQ, Isabelle Marries Dirac (IMD), and QBricks. & \citep{lewis2023formal} \\
\hline

Quantum Formal Method & QIn, a fully automated tool created to convert quantum circuits into a classical host language, Java. This allows for the analysis and verification of quantum circuits using classical formal methods techniques, enabling classical software engineers to incorporate quantum algorithms without specialized knowledge. & \citep{Klamroth2023} \\
\hline

Benchmark Datasets & QBugs, a benchmark developed to enhance reproducibility in quantum computing. It encompasses replicable bugs identified in quantum algorithms and provides supporting infrastructure for controlled experiments in quantum software testing and debugging. & \citep{Campos9474565} \\
\hline

Quantum Code Generation & A method demonstrating how to generate Python and Qiskit code from intricate UML models, incorporating essential components of higher-level classical-quantum software design. This method effectively combines current quantum code with classical code. & \citep{10313607} \\
\hline

Requirement Engineering & A study exploring the differences between quantum software and its classical counterpart in requirements engineering. It adapts traditional requirements categories to the unique context of quantum software development and considers the impact on various steps in the requirements process. & \citep{Yue10313750} \\
\hline

Refactoring & The QSharp Refactoring Tool (QRT), created to facilitate the practical restructuring of quantum programs in Q\#. Unlike tools for conventional code structures, QRT employs a program dependence graph (PDG) to depict quantum programs, enabling a comprehensive array of refactoring operations. & \citep{Zhao10313613} \\
\hline  

Reverse Engineering & Dynamic analysis to reverse Hamiltonian expressions from D-Wave quantum annealing programs, converting these expressions into a Knowledge Discovery Metamodel (KDM), allowing for smooth integration of classical and quantum software. & \citep{Pérez9860240} \\
\hline

\end{longtable}

The current literature on QSE indicates that a range of software engineering tools has been created to enhance QSE practices. Although these tools are essential for progressing the field, many are still in the developmental phase. We have discussed their strengths and weaknesses to give a clearer perspective on their functionality.

\begin{itemize}
	\item Bugs4Q excels in reproducibility with manually validated bugs but is limited to Qiskit and scalability.
	\item  Q-Diff automates bug-fix pattern recognition using ASTs for Qiskit, but may miss complex quantum-specific errors. 
	\item Quito automates testing with coverage criteria and mutation analysis, but does not address quantum-specific issues like noise.
	\item  QuSBT generates exhaustive test suites for Qiskit using a genetic algorithm, but struggles with large-scale systems. 
	\item QSharpCheck provides precise property-based testing for Q\#, but its resource intensity, scalability concerns, and dependence on Q\# are important limitations.
	\item  QMutPy automates mutation testing for Qiskit, but its scalability issues, reliance on simulations, and lack of support for other quantum frameworks are notable drawbacks.
	
	\item QIn allows classical engineers to work with quantum algorithms through Java conversion, automating the process and integrating with classical systems. However, it may compromise quantum-specific properties and face scalability issues.
	
	\item While the QBugs  excels in reproducibility, infrastructure, and multi-language support, its development and maintenance challenges, along with the difficulty of reproducing quantum bugs, present key weaknesses.
	
	\item  Q-UML builds on UML rather than developing an entirely new language, but  it may face challenges in capturing the full complexity of quantum computing and scaling to larger systems.
	
	\item The advanced refactoring features of QRT make it a valuable asset for quantum software, particularly in improving code quality. However, its limited language support and scalability concerns present potential challenges for large-scale quantum software projects.
\end{itemize}

\subsection{A case study on Q-UML for quantum software model}

UML is insufficient for quantum computing because it was initially designed for classical software systems. Quantum computing introduces unique characteristics that traditional UML cannot effectively capture, such as the probabilistic, interdependent, and non-deterministic behavior of qubits. In addition, quantum computing requires representations for quantum-specific operations, state evolutions, and measurement-driven logic, along with unique synchronization and error mitigation requirements in hybrid quantum-classical systems.  To address these challenges, quantum-specific modeling languages such as Q-UML have been developed, offering more precise and expressive tools for modeling quantum systems. This case study examines the work presented in \cite{perez2020towards}, where a minimal set of UML extensions was proposed to facilitate the effective modeling of quantum software. The paper initially establishes five fundamental design principles that any Q-UML should adhere to. These principles are described below:

\begin{itemize}

\item Quantum Classes: When a software module utilizes quantum information, whether within its internal state or as part of its interface, this must be clearly documented in the design specifications.

\item Quantum Elements: All interface elements of a module (e.g., public functions/methods, public variables) and internal state variables must be categorized as either classical or quantum and clearly labeled.
\begin{itemize}
     \item Quantum Variables: Each variable must be identified as either classical or quantum. If the model involves data types, the variables should specify the classical data types (e.g., integer, string) or quantum data types (e.g., qubit, qubit array, quantum graph state).  

     \item Quantum operations: Operations should clearly label both input and output as classical or quantum and indicate if the operation itself is quantum-based. 
\end{itemize}

\item Quantum Supremacy: A module containing at least one quantum element is considered a quantum software module; otherwise, it is a classical module. These should be clearly labeled.

\item  Quantum Aggregation: A module composed of one or more quantum modules is itself a quantum module and must be labeled as such.

\item  Quantum Communication: Quantum and classical modules can communicate if their interfaces are compatible, meaning the quantum module has classical inputs or outputs that can connect with the classical module.

\end{itemize}

As an extension of UML, Q-UML encompasses all the functionalities of UML while additionally accounting for quantum information. It ensures that any instance of quantum information—whether stored, transmitted, or processed—is explicitly marked. Both structural and behavioral diagrams are supported, including use case, class, object, state machine, sequence, and activity diagrams. In QUML, quantum information can be represented either visually or textually, with text-based elements using bold font to differentiate them from classical data, while visual diagrams utilize double lines to set quantum information apart.

In Q-UML, all static (e.g., classes) and dynamic structures (e.g., processes) are initially considered classical. A structure is designated as quantum only if it directly stores quantum information or if any of its components (like variables or aggregated classes) involve quantum information. Similarly, a dynamic process is considered quantum if it performs quantum information processing or if any of its sub-processes engage in quantum information processing.

\begin{figure}[!htp]
    \centering
    \includegraphics[width=0.5\linewidth]{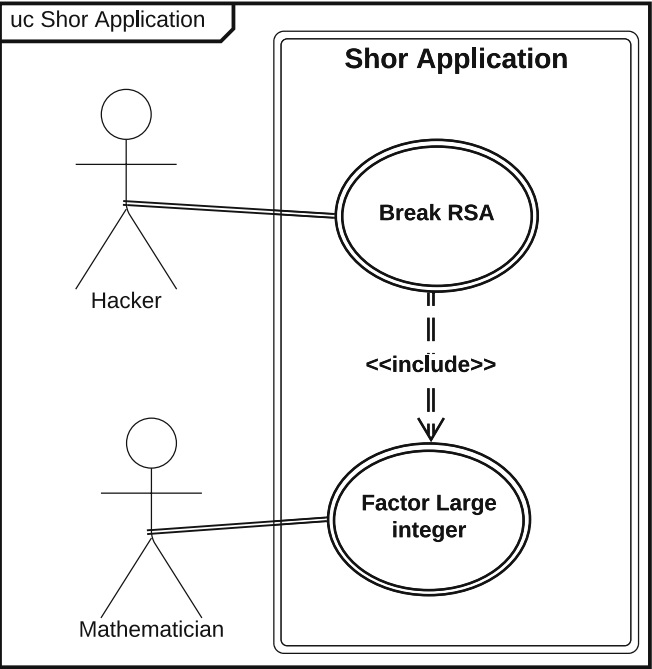}
    \caption{An Q-UML Use-case diagram for Shor Application }
    \label{fig:qml_usecase}
\end{figure}

This paper demonstrates Q-UML using the Shor Application, a quantum software implementation of Shor's algorithm. In practice, the system finds applications in cybersecurity, as well as in number theory, combinatorics, and optimization. Initially, a Q-UML use case diagram for this Shor Application is presented.  Figure \ref{fig:qml_usecase} shows the Q-UML Use-case diagrams for the Shor application, where use cases requiring quantum resources are distinctly marked with a double line.

\begin{figure}[!htp]
    \centering
    \includegraphics[width=0.75\linewidth]{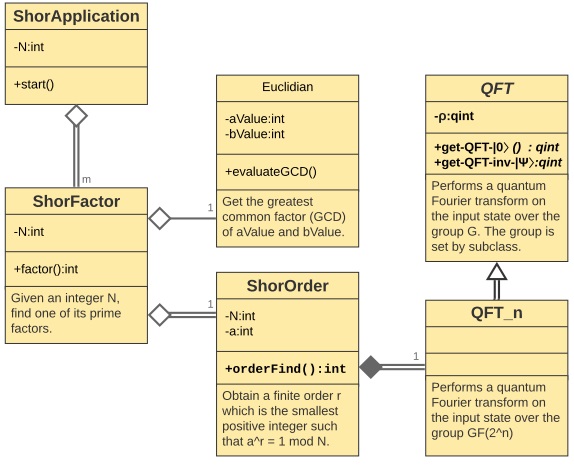}
    \caption{An Q-UML  class  diagram for Shor Application }
    \label{fig:qml_class}
\end{figure}

Figure \ref{fig:qml_class} presents a Q-UML class diagram illustrating the Shor application and demonstrating the application of the specified rules to both class and object diagrams. The Shor application consists of six classes, five of which—ShorApplication, ShorFactor, ShorOrder, and QFT\_n—are quantum in nature, while the Euclidean class is the only classical class in the configuration.

To represent quantum information in UML diagrams, bold formatting is used for quantum elements. Quantum attributes are indicated by bold names. For methods, bold inputs indicate quantum inputs, and bold datatypes indicate quantum outputs. If datatypes are omitted, they're assumed classical. If a class or object has quantum attributes or methods, its name is also bold, indicating it's a quantum class or object. Quantum class relationships are shown with double lines. In inheritance, a quantum superclass makes the subclass and relationship quantum. In aggregation/composition, a quantum component makes the container and relationship quantum. Associations require quantum classes to have a classical interface for interacting with classical ones.

Besides this, sequence diagrams, activity diagrams, and state diagrams for the Shor application can be presented, as shown in the original paper. It is noted that this minimal extension of classical UML to Quantum UML can be effective for further QSE development. This extension of the model is complete and capable of modeling any quantum software application.

\subsection{Research Agenda for QSE}

After reviewing case studies and tools related to QSE, it is evident  that  the field is still in its early stages of development, and many of the associated tasks are progressing.   This observation highlights the need for more research and development to strengthen both the practical use and the theoretical foundation of QSE. Hence, providing a strategic framework to overcome the current challenges in the QSE field, we have considered several research agenda along  with research questions, objectives, and probable methodologies, all of which are presented in  Table \ref{tab:agenda}.

\begin{longtable} {|p{2cm}|p{3.5cm}|p{3.25cm}|p{3.25cm}|}
	\caption{Some Research Agenda for Advancing QSE}
	\label{tab:agenda} \\
	
	\hline
		\textbf{Agenda} & \textbf{Research Questions} & \textbf{Objectives} & \textbf{Methodologies} \\
	\hline
	\endfirsthead
	
	\multicolumn{3}{c}{{\bfseries \tablename\ \thetable{} -- continued from previous page}} \\
	\hline
		\textbf{Agenda} & \textbf{Research Questions} & \textbf{Objectives} & \textbf{Methodologies} \\
	\hline
	\endhead
	
	\hline \multicolumn{3}{|r|}{{Continued on next page}} \\ 
	\hline
	\endfoot
	
	\hline 
	\endlastfoot

\hline Developing Higher-Level Abstractions for Quantum Software
& How can we create higher-level abstractions for quantum software that align with classical software engineering practices? & - Develop modeling techniques that incorporate quantum principles.\newline - Create quantum-specific design patterns and architectures.\newline - Establish hybrid quantum-classical software modeling approaches.  & - Compare classical and quantum software development methods. \newline - Develop domain-specific languages for quantum computing. \newline - Create and evaluate quantum software modeling tools.
\\

\hline Quantum Software Development Tools and Ecosystems & What are the key requirements for developing quantum software development tools? \newline How can classical tools be adapted for quantum software development? & - Create or improve quantum software development tools. \newline - Design integrated ecosystems for quantum software development. & - Build quantum IDEs, debuggers, and simulators. \newline -  Conduct studies to improve development tools. \newline -  Partner with hardware manufacturers to create integrated ecosystems. \\

\hline Quantum Software Testing and Verification & What new techniques are needed for testing and verifying quantum software? & - Develop quantum-specific testing methods. \newline
	- Create debugging and error correction tools. \newline
	- Establish formal verification for quantum algorithms. & - Adapt classical testing techniques for quantum systems.
\newline - Develop quantum circuit simulators for testing.
\newline - Create formal verification frameworks for quantum programs. \\

\hline Quantum Software Maintenance and Lifecycle & What is the lifecycle of quantum software, and how can it be managed efficiently? \newline How can software evolve with emerging hardware and algorithms? & - Establish quantum software lifecycle practices. \newline - Investigate challenges of maintaining quantum software over time. & - Gather insights from quantum software engineers. \newline - Develop quantum-specific lifecycle models. \newline -  Study long-term challenges in quantum software maintenance. \\
	
\hline Workforce Development and Training & How can classical software engineers be effectively trained for quantum software engineering? & - Define training pathways for classical-to-quantum transition.\newline -  Develop QSE methodologies to ease the learning curve.\newline -  Create educational resources bridging quantum mechanics and software engineering. & - Develop curriculum for QSE.\newline -  Analyze case studies of successful transitions.\newline -  Develop and test quantum programming tutorials and workshops.\\

\hline

\end{longtable}

\section{ Opportunities and Challenges}

Despite significant progress in quantum computing, the development of practical large-scale quantum systems remains challenging due to hardware limitations and the complexity of implementing quantum algorithms at scale \citep{AJAGEKAR201976}. Qubits are highly sensitive to environmental disturbances, leading to decoherence and making it difficult to perform complex computations before quantum states are lost. Scaling quantum systems to hundreds or thousands of qubits while maintaining stability, coherence, and low error rates is a major hurdle, as current systems are small and prone to noise-induced errors.
Effective error correction is crucial for ensuring reliable quantum computations, but it demands additional qubits, which significantly complicates the scalability of quantum systems \citep{Bravyi22}.  Additionally, developing quantum algorithms that efficiently utilize near-term systems is challenging, as they often require substantial resources, including a large number of qubits and gates. The complexity of gate operations, state preparation, and readout presents additional challenges \citep{Volya10413647}. Moreover, simulating large-scale quantum algorithms on classical computers is both computationally expensive and time-intensive.

The following presents a brief overview of the challenges and opportunities offered by QML and quantum optimization algorithms within the context of software engineering research. Furthermore, we outline the challenges and opportunities related to QSE.

\subsection{QML in Software Engineering Research}
\subsubsection{Challenges:}
\begin{itemize}

 \item  Quantum machines are costly and less readily available than classical computers. Additionally, challenges such as decoherence and qubit errors present significant obstacles, complicating the advancement and integration of efficient QML algorithms. This, in turn, makes it difficult to fully harness the potential of quantum computing  for practical applications.

 \item In software engineering research, high-dimensional datasets are commonly used. However, employing a large volume of data in QML for software engineering tasks is difficult due to the limited number of available qubits. Although many SDKs offer simulations that can be executed on classical computers, these simulations only partially capture the capabilities of a real quantum machine. This presents a noteworthy challenge when using QML to process large datasets.

 \item A major challenge of current QML is the prevalence of hybrid models. These models utilize both classical and quantum computing resources but often fail to harness the potential of quantum computing fully. Creating a purely quantum-based machine learning model, especially for linear tasks, remains a subject of active research and advancement.

\end{itemize}

\subsubsection{Opportunities}

\begin{itemize}
\item In the field of software engineering research, where classical machine learning techniques have traditionally been used, there is an opportunity to explore the integration of QML algorithms. While significant progress has been made in classical machine learning applications, the use of QML algorithms in software engineering still needs to be explored. Various domains within software engineering, such as bug prediction and code smell detection, could benefit from the potential capabilities of QML.

\item To build applications and perform analysis using QML algorithms, numerous SDKs and APIs have been developed to facilitate QML. These tools enable developers and practitioners to build models quickly and efficiently address various software engineering problems. The availability of these SDKs and APIs accelerates the integration of QML into software engineering, enabling quick and efficient problem-solving in the field.

\item Dealing with high-dimensional data is common in many software engineering tasks. QML algorithms offer improved speed and efficiency compared to classical Machine Learning (ML) by leveraging quantum parallelism, allowing for simultaneous computation across multiple dimensions. This characteristic makes QML a promising solution for handling complex datasets in software engineering applications.
\end{itemize}

\subsection{Quantum Search and Optimization Algorithms  in Software Engineering Research}
\subsubsection{Challenges}
\begin{itemize}

    \item To effectively utilize quantum optimization algorithms in software engineering-related problem-solving, it is essential to transform the problem into an appropriate encoding, such as QUBO or a related optimization problem. This transformation process can be intricate and challenging. As not all software engineering problems are suitable for quantum optimization algorithms, it is crucial to identify the specific types of software engineering problems that can benefit most from these approaches.

\item  Understanding and implementing quantum optimization algorithms through quantum machines for complex software engineering problems requires a deep understanding of quantum theory, advanced mathematics, and low-level details of quantum machines. 

\item Quantum optimization algorithms such as QAOA, VQE, or Quantum Annealing  do not guarantee the global best solution. However, they are designed to find near-optimal solutions more quickly and efficiently than traditional classical optimization approaches.

\item The quality of the final solution for a software engineering problem is not solely dependent on the quantum optimization algorithm used. It also significantly relies on the problem-related objective function and the capabilities of the quantum computing machine where it is run.

\item  The performance of quantum optimization algorithms and the ability to solve large-scale software engineering problems are currently hindered by the limited qubit count and high error rates associated with existing quantum computers. Moreover, access to quantum hardware and simulators for research is often restricted and costly. The absence of standardized quantum programming languages and development tools also poses significant challenges in implementing quantum optimization algorithms.

\end{itemize}

\subsubsection{Opportunities}

\begin{itemize}

\item Quantum optimization methods have found applications across diverse domains, ranging from finance and sustainability to engineering design, encompassing areas like materials design and topology optimization. However, within software engineering, the exploration of quantum optimization still needs to be improved. Since search-based software engineering often involves dealing with optimization problems, adopting quantum optimization could be a beneficial approach.

\item Due to quantum properties, quantum optimization algorithms support massive parallel searching, allowing quicker identification of the optimal solution. When solving large-scale optimization problems related to software engineering, these quantum algorithms can be significantly faster than their classical counterparts.

\item Major quantum SDKs offer high-level programming interfaces, libraries with APIs for connecting to quantum machines, and simulators to implement major quantum optimization algorithms such as QAOA  and VQE. By leveraging these resources, various software engineering problems can be formulated and solved using quantum optimization algorithms, albeit on a limited scale.
\end{itemize}

\subsection{Challenges in QSE}

\subsubsection{Challenges}
\begin{itemize}

 \item One of the significant challenges is understanding quantum computing to perform software engineering tasks in quantum-based applications.
Aoun et al. \citep{aoun9609196} noted from QSE-related questions on Stack Exchange forums that areas of challenge in QSE differ from traditional software engineering, including the need to explain the theory behind quantum computing code and interpret quantum program outputs.  Therefore, bridging the knowledge gap between quantum and classical computing is a prerequisite for better QSE practice.

 \item  Nowadays, quantum-based applications are often hybrid, consisting of both quantum and classical components. Performing QSE in these cases involves using both quantum and classical software development processes, which often results in integration challenges.

\item To identify the challenges in QSE, \cite{Akbar2023}  conducted a study where they identified 22 challenging factors in the QSE process and algorithms. They also ranked them based on the importance of each challenge. The top five identified challenges are limited software libraries, maintenance complexity, complex programming, lack of training and workshops, and data encoding issues.

\item Another major limitation of QSE is the need for a well-established body of knowledge. The methods used to test quantum algorithms and protocols are still fundamental. While the primary goals of SE involve increasing abstraction levels, quantum algorithms are primarily based on circuit representations, which means they do not take advantage of higher-level abstractions.



\end{itemize}

\subsubsection{Opportunities}

\begin{itemize}

\item The tools and techniques currently developed for quantum software engineers are primarily at a preliminary stage. Therefore, there is a pressing need to build more advanced tools and libraries tailored for quantum software development—these range from quantum requirements to quantum testing and debugging. The introduction of these new paradigms and techniques has the potential to improve the software development process significantly.

\item Quantum computing is still in its early stages of development. As quantum computing advances, new quantum applications will be developed. This indicates that a properly structured QSE approach will aid in developing well-structured quantum applications, facilitating innovation and domain-specific quantum software solutions. Besides, developing high-level abstractions and frameworks makes quantum programming more accessible to software engineers without extensive knowledge of quantum theory and circuit details.

\item The integration of QSE with classical software engineering will aid in constructing large-scale hybrid quantum applications. The development of these hybrid applications necessitates the creation of appropriate quantum software architectures, which, in turn, presents opportunities for innovation in the field of software architecture.

\end{itemize}

\section{Discussion} In this section, we have answered the research question we raised in the introduction.
\begin{itemize}

\item \textbf{Answer to RQ1:} Our comprehensive literature review indicates that QNN and QSVM models are predominantly used to detect software vulnerabilities. In addition, QSVM models have found applications in detecting code smells. This field has seen limited research. However, there are other aspects of software engineering where QML could prove beneficial, such as predicting bugs, detecting spam, analyzing sentiment, and classifying code. Considering the successful implementation of QML in various real-world situations, different quantum computing models could be applied to software engineering tasks. We have presented a generic model that uses QML algorithms to address issues related to software engineering. This model could serve as a guide for researchers in creating their predictive models.

\item \textbf{Answer to RQ2:} Our literature review reveals that Grover’s algorithm has been employed in software testing applications. In software engineering, problems like code clone detection and software verification are often converted into graph-based combinatorial optimization problems. These challenges are then addressed using quantum optimization techniques, including QAOA and Quantum Annealing. Despite the success of quantum optimization in various computationally complex real-world problems, its application in the software domain remains relatively unexplored. We have presented a general approach that allows researchers to apply QAOA and Quantum Annealing to software engineering problems, potentially broadening the scope of quantum computing applications in this area.

\item \textbf{Answer to RQ3}: Our review of QSE indicates that while QSE tools were quite limited prior to 2020, there has been a significant increase in the development of new tools in the field. This surge is driven by the growing popularity of quantum-based applications, which has led to increased interest and focus on QSE. We observed that most tools and techniques presented in the literature are related to quantum software testing and debugging, followed by empirical QSE and quantum formal methods. Although classical software engineering principles can serve as a foundation for developing many QSE tools, the distinctive characteristics of quantum computing require unique approaches in QSE.

\item \textbf{Answer to RQ4}: We have addressed this research question in Section 6, where we discuss several opportunities and challenges in QSE, as well as quantum-assisted software engineering research.

\end{itemize}

\section{Threats to Validity}

In this section, we have pinpointed several factors that could impact the validity of our review process. In our article search, we carefully constructed several search queries. However, they might not have been all-encompassing. To counteract this risk, we attempted to manually search major quantum computing and software engineering-related conferences and proceedings to include papers initially missed by our search string. Before the final selection process, we carefully considered the relevance of the documents for inclusion in specific groups. One limitation of this study is that we did not consider unpublished papers, significantly pre-print articles, technical papers, and PhD or Master’s theses. This selection criteria might have resulted in a bias in article selection. However, we concentrated on peer-reviewed and published articles in quality index datasets, thereby avoiding grey literature. To ensure that the review could adequately cover the domain of quantum computing and software engineering, we also repeated the paper selection process with the assistance of other co-authors.

\section{Conclusions}

The main goal of the review article is to investigate and understand how the fields of quantum computing and software engineering intersect. Therefore, we structure our review process into two main areas. The first is how quantum computing can expedite the software engineering process, and the second is the current state of QSE. Initially, we overview the fundamentals of quantum computing, including its distinction from classical computing and the inherent quantum gates and circuits for quantum systems, quantum algorithms, and SDKs. To assess the viability of quantum computing for software engineering research, we have examined the capabilities of QML and Quantum Search and Optimization algorithms. We found that limited work has been conducted on quantum computing for software engineering tasks. We also present generic workflows demonstrating how quantum computers could be employed for various software engineering tasks. It is apparent that with the advancement of quantum applications in several fields, more quantum algorithms and quantum programming have been developed, which necessitates appropriate QSE. We examine the progress in QSE, describing the attempts to develop tools and techniques to enhance the QSE domain.


Our investigation shows that the potential of quantum computing, including QML and quantum optimization algorithms for addressing software engineering tasks, is immense. Until now, not much work has been done in this area, indicating several open research directions. Furthermore, the QSE tools that have been developed to support quantum-based applications are not adequately meeting the requirements or demands of the field. This situation demands more robust and versatile tools to support the development and deployment of quantum-based applications effectively. It has also been noted that substantial challenges persist, including the lack of large-scale quantum computers, the noise associated with quantum computing, high expenses, and limited accessibility compared to traditional computers. Despite these challenges, we remain optimistic that the insights provided in this review will aid researchers and practitioners in understanding the potential of quantum computing for software engineering research and the implications of QSE for the development of quantum applications.

\section*{Declarations}

\begin{itemize}
\item Funding: This research is supported in part by the Natural Sciences and Engineering Research Council of Canada (NSERC) Discovery Grants program, and by the industry-stream NSERC CREATE in Software Analytics Research (SOAR). 
\item Conflict of interest: The authors declare that they have no conflict of interest.
\item Competing interests: The authors declare no competing interests.
\item Ethics approval and consent to participate: Not applicable.
\item Consent for publication: Not applicable.
\item Data availability: The datasets supporting the results of this article are available upon request.
\item Materials availability: Not applicable.
\item Code availability: Not applicable.
\item Author Contributions: All authors contributed to the conception and design of the review. Ashis Kumar Mandal handled the material preparation, data collection, and analysis. He also drafted the initial manuscript, which was reviewed and revised by Md Nadim. Chanchal K. Roy, Banani Roy, and Kevin A. Schneider provided supervision and project administration. All authors read and approved the final manuscript.
\end{itemize}

\bibliography{sn-bibliography}

\end{document}